\documentclass[a4paper,12pt]{article}
\usepackage{graphicx, array, blindtext}
\usepackage{color}
\usepackage{color, colortbl}
\usepackage{amsmath}
\usepackage{latexsym}
\usepackage{rotating}
\usepackage{setspace}
\usepackage[margin=1in]{geometry}
\usepackage{times}
\usepackage{txfonts}
\usepackage{mathptmx}
\usepackage{rotating}
\usepackage{subfloat}
\usepackage{subfig}
\usepackage{authblk}
\usepackage{multirow}
\usepackage{listings}
\renewcommand{\arraystretch}{1}

\usepackage{cite}

\makeatletter
\renewcommand{\@biblabel}[1]{\quad#1.}
\makeatother

\renewcommand{\arraystretch}{1}

\definecolor{LightCyan}{rgb}{0.88,1,1}

\usepackage{color} 
\definecolor{mygreen}{RGB}{28,172,0} 
\definecolor{mylilas}{RGB}{170,55,241}
\title{\huge{A mechanistic model quantifies artemisinin-induced parasite growth retardation in blood-stage \emph{Plasmodium falciparum} infection}}

\author[1]{Pengxing Cao}
\author[2]{Nectarios Klonis}
\author[3]{Sophie Zaloumis}
\author[4]{David S. Khoury}
\author[4]{Deborah Cromer}
\author[4]{Miles P. Davenport}
\author[2]{Leann Tilley}
\author[3]{Julie A. Simpson}
\author[1,3,5]{James M. McCaw\thanks{Correspondence: jamesm@unimelb.edu.au}}
\affil[1]{School of Mathematics and Statistics, The University of Melbourne, Melbourne, Australia.}
\affil[2]{Department of Biochemistry and Molecular Biology, Bio21 Molecular Science and Biotechnology Institute, University of Melbourne, Melbourne, Australia.}
\affil[3]{Centre for Epidemiology and Biostatistics, Melbourne School of Population and Global Health, The University of Melbourne, Melbourne, Australia.}
\affil[4]{Infection Analytics Program, Kirby Institute, UNSW Australia, Kensington, New South Wales, Australia.}
\affil[5]{Modelling and Simulation, Infection and Immunity Theme, Murdoch Childrens Research Institute, The Royal Children's Hospital, Parkville, Victoria, Australia.}

\begin{document}

\date{}
\maketitle
\vspace{1.5cm}

 \thispagestyle{empty}

\newpage

\section*{Abstract}

Falciparum malaria is a major parasitic disease causing widespread morbidity and mortality globally. Artemisinin derivatives---the most effective and widely-used antimalarials that have helped reduce the burden of malaria by 60\% in some areas over the past decade---have recently been found to induce growth retardation of blood-stage \emph{Plasmodium falciparum} when applied at clinically relevant concentrations. To date, no model has been designed to quantify the growth retardation effect and to predict the influence of this property on \emph{in vivo} parasite killing. Here we introduce a mechanistic model of parasite growth from the ring to trophozoite stage of the parasite's life cycle, and by modelling the level of staining with an RNA-binding dye, we demonstrate that the model is able to reproduce fluorescence distribution data from \emph{in vitro} experiments using the laboratory 3D7 strain. We quantify the dependence of growth retardation on drug concentration and demonstrate the model's utility as a platform to propose experimentally-testable mechanisms of growth retardation. Furthermore we illustrate that a drug-induced delay in growth may significantly influence \emph{in vivo} parasite dynamics, demonstrating the importance of considering growth retardation in the design of optimal artemisinin-based dosing regimens.

%

\newpage

\section*{Introduction}

\emph{Plasmodium falciparum} malaria is a major parasitic disease which causes severe morbidity and mortality in approximately half a million people annually \cite{WHOrep2015}. Artemisinin (ART) and its derivatives (e.g.\ artesunate, dihydroartemisinin and artemether), used in combination with partner drugs, provide front-line protection, and have been responsible for dramatic reductions in disease burden over the past few decades \cite{WHOrep2015}. Despite their clinical and public health effectiveness, the emergence of ART resistance and lack of alternative treatments places current control programs at risk \cite{Dondorpetal2009,Phyoetal2012,Arieyetal2014,Ashleyetal2014}. Development of a comprehensive understanding of ART's mechanism of action and associated effects on infected red blood cells (iRBCs) is therefore critical for development of optimised ART-based treatment regimens and maintenance of control program impact \cite{Simpsonetal2014}.

Recent \emph{in vitro} experiments \cite{Klonisetal2013,Dogovskietal2015,Yangetal2016}, combined with advances in pharmacokinetic--pharmacodynamic (PK--PD) modelling \cite{Caoetal2016}, have established a platform to probe the parasite's temporal response to antimalarial drugs. The key experimental advance underlying these \emph{in vitro} studies was the application of short drug pulses, which enabled fine-scale measurement of the killing effect of drug \cite{Caoetal2016}. A normal life cycle of an iRBC for \emph{P. falciparum} is approximately 48 hours and is classified based on morphological appearance into three main stages: the ring stage (approximately 0--26 hours post infection (h p.i.)), trophozoite stage (approximately 27--38 h p.i.) and schizont stage (approximately 39--48 h p.i.). Upon rupture at approximately 48 h p.i., iRBCs release merozoites, 8--12 of which successfully invade susceptible RBCs to initiate a new round of infection \cite{Simpsonetal2002,Dietzetal2006,Zaloumisetal2012}. Dogovski \emph{et al.} demonstrated that a short pulse of ART (or dihydroartemisinin (DHA)) can induce growth retardation, prolonging the 48 hour life cycle \cite{Dogovskietal2015}. Importantly, they found that growth retardation did not stop parasite growth entirely and was thus considered to be distinct from parasite dormancy which ``freezes'' parasites for days to weeks \cite{Teuscheretal2010,Coddetal2011}. 

Experimental identification of drug-induced growth retardation raises two questions: 1) By how much is the life cycle of a parasite prolonged in response to a short drug exposure pulse?; and 2) How influential is growth retardation when considering \emph{in vivo} parasite killing using PK--PD models? These two questions are important because we expect any drug-mediated variation in the duration of one or more life stages to impact on the efficacy of the drug (given the well-established finding that drug can exert stage-specific killing effect to parasites \cite{Saralambaetal2011,Klonisetal2013,Witkowskietal2013,Caoetal2016}.) The alteration in drug efficacy may be significant if the prolonged stage(s) covered by the drug pulse exhibit very distinct killing effects. Quantification of growth retardation and assessment of its potential effect on \emph{in vivo} parasite killing is therefore important in guiding further experimental investigations into drug activity and strategies for optimising ART-based combination therapies.

To the best of our knowledge, no model has yet been designed to quantify the growth retardation effect and predict its influence on \emph{in vivo} parasite killing. Here we construct a mechanistic model of parasite growth to explain drug-induced growth retardation. We model the relationship between parasite age and the fluorescence intensity of a parasite's RNA/DNA-binding dye. Through application to fluorescence data from \emph{in vitro} experiments using the laboratory 3D7 strain by Dogovski \emph{et al.}\cite{Dogovskietal2015}, we provide the first quantification of growth retardation. We assess its dependence on drug (ART and DHA) concentration. We also suggest specific alternative hypotheses for the mechanism of drug-induced growth retardation and demonstrate that those mechanisms can be reliably identified in future experimental studies. Finally, by incorporating growth retardation into a PK--PD modelling framework, we simulate \emph{in vivo} parasite killing upon exposure to a single dose of artesunate and show how growth retardation may manifest as a phenotypical indication of drug-resistance.

\section*{Materials and Methods}

\subsection*{Experiment and data}

We first summarise the \emph{in vitro} experiment in which parasite growth retardation was identified. We provide sufficient information for the purposes of model development and evaluation. For full details on the experimental implementation we refer the reader to the original publication \cite{Dogovskietal2015}. 

\begin{figure}[ht!]
\centering
\includegraphics[scale=1]{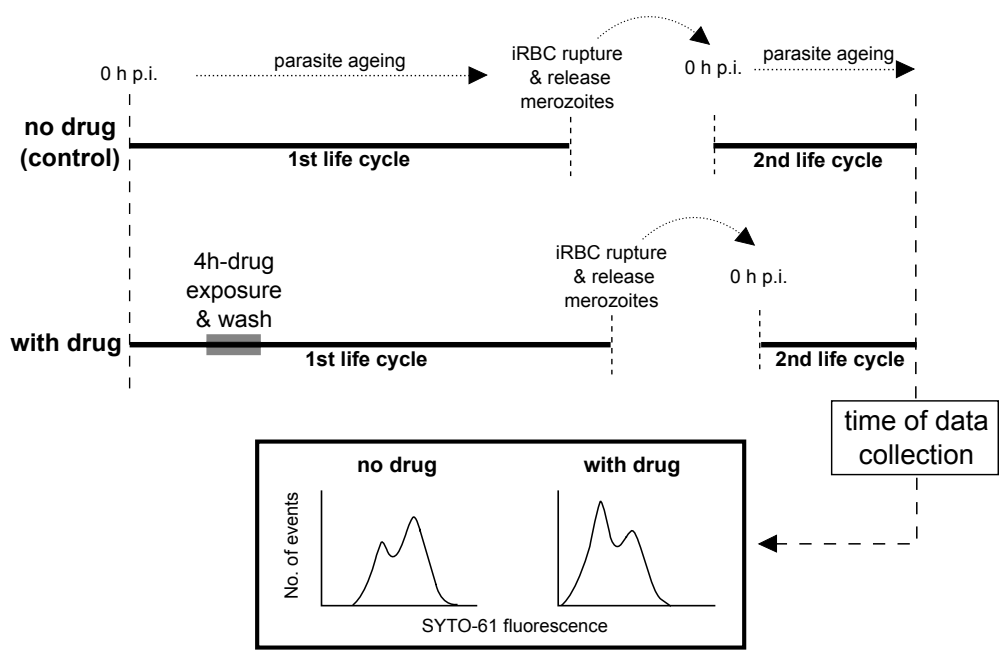}
\caption*{Figure 1: \small{Schematic diagram showing the experimental process. Cultured synchronised parasites (3D7 strain) either experience intraerythrocytic development without treatment (no drug) or were exposed to a 4h drug pulse at approximately 6 h p.i. (with drug). Upon rupture of infected RBCs, released merozoites infect susceptible RBCs to initiate the second life cycle. At the time of data collection in the second life cycle (which was carefully adjusted for an optimal observation of the bimodal fluorescence intensity distribution), SYTO-61 fluorescence intensity data was collected and displayed using histograms (e.g. presented in the box). Trophozoites exhibit higher fluorescence intensities than rings. It follows that the left mode of the distribution primarily represents rings while the right one represents trophozoites. Any change in the distribution indicates a possible drug-induced age retardation effect. Further details are provided in the \emph{Materials and Methods}.}}
\end{figure}

Fig.\ 1 presents the experimental process. A culture containing tightly synchronised rings (3D7 strain; over 80\% of the population within a one-hour age window) with an average age of 6 h p.i.\ was equally divided into a number of small cultures, each of which was treated with a different concentration of drug (ART or DHA) for 4 hours. Two cultures unexposed to drug acted as the control. The cultures were stained with SYTO-61 and examined by flow cytometry \cite{Fuetal2010}. SYTO-61 is a nucleic acid stain which stains both DNA and RNA and allows the distinction of infected RBC from uninfected RBC, as well as distinguishing between parasite-infected RBC of different ages due to an increase in nucleic acid content as the parasite ages. SYTO-61 signals from all cultures were collected simultaneously at approximately 72 hours post drug administration (indicated in Fig.\ 1), corresponding to the period in which parasites transition from the ring to trophozoite stage (during the second cycle). As trophozoites express significantly more nucleic acid, this period of transition exhibited a bimodal SYTO-61 fluorescence distribution. Quantitative analysis of this bimodal distribution is the key to our approach as it enables us to identify the relative populations of parasites in the two successive life stages \cite{Dogovskietal2015}. As indicated in Fig.\ 1, if the administration of drug slows parasite growth then the first life cycle would be prolonged and the second life cycle would start later. In consequence, we would observe an increase in the ring population (i.e. the mode with the lower fluorescence) and a corresponding decrease in the trophozoite population (i.e. the mode with the higher fluorescence) in the SYTO-61 fluorescence histogram. Note that the underlying processes of iRBC rupture and merozoite release and re-infection were not observable in the experiment.

\begin{figure}[ht!]
\centering
\includegraphics[scale=1]{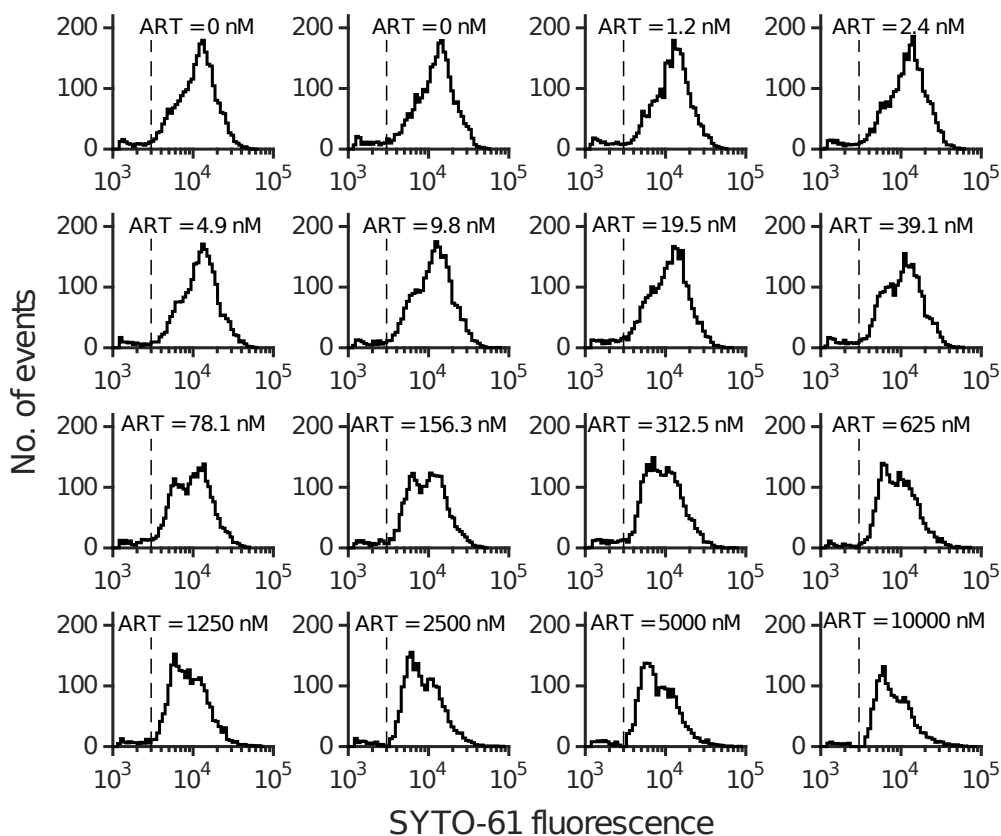}
\caption*{Figure 2: \small{SYTO-61 fluorescence frequency histograms with various ART concentrations (note that two cultures with 0 nM ART were measured). For each panel, samples with fluorescence less than 3000 (indicated by the dashed lines) were taken to include fluorescence signals from uninfected RBC and were thus excluded from the model fitting process. Note that the histograms are corrected by removing the unviable population (see \emph{Materials and Methods} for details).}}
\end{figure}

The SYTO-61 fluorescence histograms often contain a small population of dead or dormant parasites (due to drug activity or otherwise) that are not involved in the second life cycle. To account for this non-viable population, a \emph{background experiment} in which supermaximal drug concentration ($> 10 \times$ the 50\% Lethal dose (3 days)) was applied for over 48 hours was also performed \cite{Dogovskietal2015}. The high drug concentration and long exposure time guaranteed that all parasites became non-viable. Hence, denoting the SYTO-61 fluorescence frequency histogram under the \emph{background} condition by $f_b$ and the SYTO-61 fluorescence frequency histogram under a 4h drug pulse by $f$, the corrected SYTO-61 fluorescence frequency histogram, $f_c$, is given by
\begin{equation}
f_c = f - (1-V)f_b,
\end{equation}
where $V$ represents the viability (the fraction of parasites entering the second life cycle; see \cite{Klonisetal2013,Dogovskietal2015} for details). 

The corrected histogram data is shown in Fig.\ 2 (for various ART concentrations) and Supplementary Fig.\ S1 (for various DHA concentrations). Experiments were performed in technical replicates for each drug concentration. The histograms in Fig.\ 2 present all available SYTO-61 fluorescence intensity data. Also note that for each histogram, the samples with fluorescence intensity less than 3000 (indicated by the vertical dashed lines) were considered to include fluorescence signals from uninfected RBCs and were thus excluded in the analysis. Each histogram was generated by distributing log-transformed SYTO-61 fluorescence intensity samples (with magnitude $>$3000) into 40 equally spaced bins. Raw SYTO-61 fluorescence intensity data is provided in Dataset S1.

\subsection*{The model}

Since the \emph{in vitro} experiment measures the SYTO-61 fluorescence data in the second life cycle, our model is designed to reproduce the underlying process of parasite growth over the period of the ring-to-trophozoite transition in the second life cycle (see Fig.\ 1).  

Over the period of the ring-to-trophozoite transition, the growth of individual parasites is modelled by two sequential stages --- an ``immature" ring stage during which the ring-to-trophozoite transition cannot occur (due to incomplete cellular development) followed by a ``mature" ring stage where the ring-to-trophozoite transition is possible --- inspired by the classic model for the mammalian cell cycle \cite{Smithetal1973,Brooksetal1980}. We introduce $A_r$, the \emph{ready-for-change} age (which is assumed to be the same for all parasites). Parasites of age $a<A_r$ are, by definition, rings. For parasites of age $a>A_r$, the waiting time before entering the trophozoite stage follows a Poisson distribution with transition rate $\lambda$ (denoted as $Pois(\lambda)$).

We assume that the age distribution of viable parasites (i.e. the parasites able to asexually reproduce in the second life cycle) at the time of data collection is Gaussian ($\sim N(\mu,\sigma^2)$). This is reasonable given that the parasites are tightly synchronised and drug is unlikely to differentially kill parasites over such a tight age window (although a spreading in age distribution over time was evident \cite{Dogovskietal2015}.) 

The bimodal fluorescence intensity distributions (Fig.\ 2) suggest that both the SYTO-61 fluorescence intensity and the rate of increase in SYTO-61 fluorescence intensity is higher in  trophozoites than in rings. We model this property using a piecewise function mapping from parasite age ($a$) to SYTO-61 fluorescence intensity ($F$) as follows:
\begin{equation}
\label{eq:2}
F(a) = \left\{\def\arraystretch{1.2}%
  \begin{array}{@{}c@{\quad}l@{}}
     F_0e^{r_1a}, & a < A_{c} \\
     F_0e^{r_1A_{c}+r_2(a-A_{c})}, & a \geq A_{c}\\
  \end{array}\right.
\end{equation}
where $A_c$ indicates the age of transition (from ring to trophozoite) for a parasite, noting that each individual parasite's $A_c$ is sampled from \emph{Pois}($\lambda$). Clearly $A_c \geq A_r$ for all parasites. For each parasite, prior to the transition (i.e. $a < A_c$), the age-dependent intensity of SYTO-61 staining increases at a rate $r_1$. Following the transition the rate increases to $r_2$. We note that although $F_0$ gives the fluorescence intensity at the start of the second life cycle (i.e. age 0 h p.i.), the model is not designed to study the early intraerythrocytic life-stages of the parasites.

All model parameters are listed in Table 1. We note two important limiting cases of the model. When $A_r = 0$, the model allows rings to transition to trophozoites immediately (at 0 h p.i.), with transition rate $\lambda$. On the other hand, when $\lambda \to \infty$, all rings reaching age $A_r$ immediately undergo the transition to the trophozoite stage. Therefore, our model provides a very general framework in which to study the statistics of the ring-to-trophozoite transition and the mechanism of growth retardation.

\begin{figure}[ht!]
\centering
\includegraphics[scale=1]{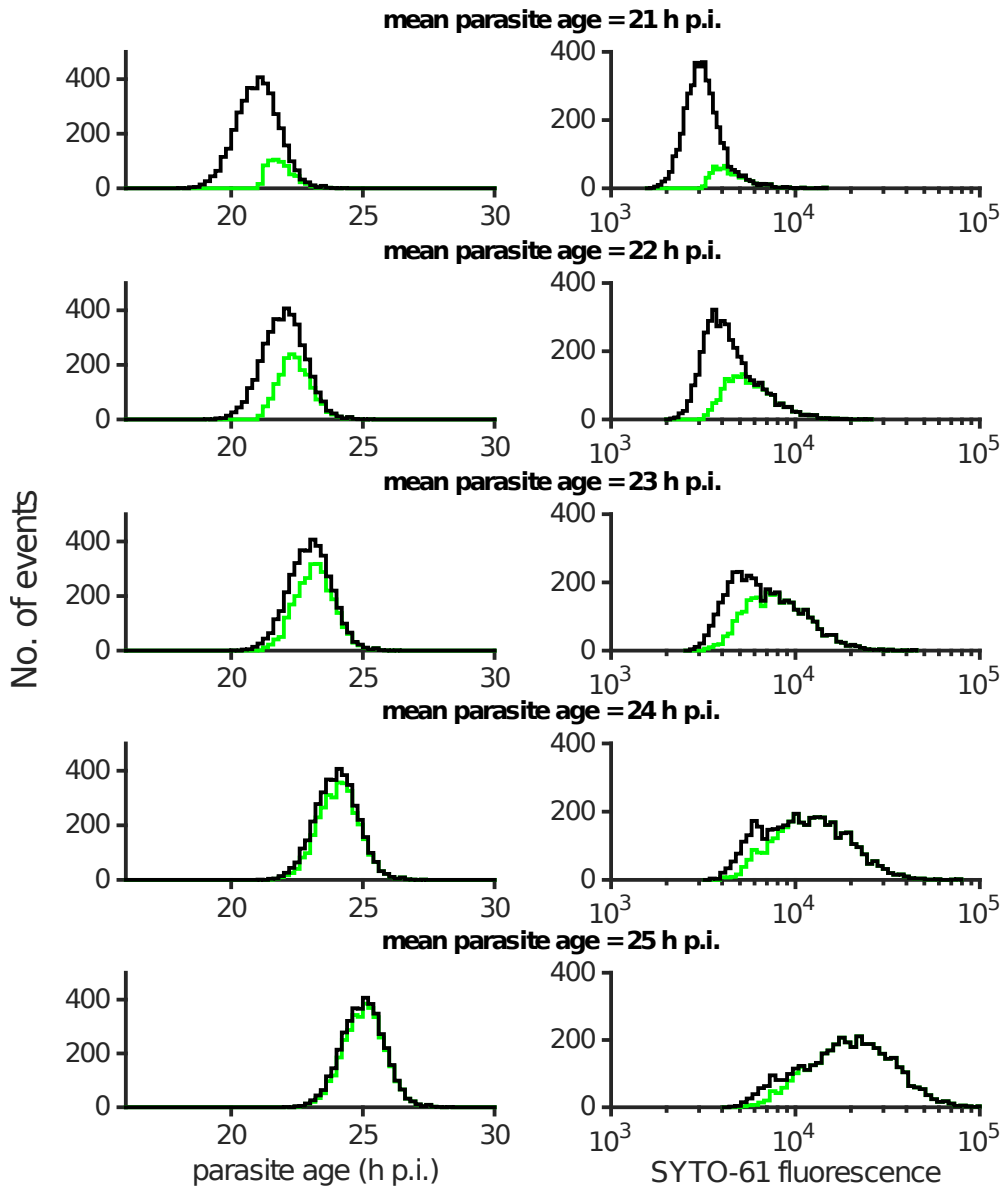}
\caption*{Figure 3: \small{A single stochastic realisation of the model showing the evolution of both the age distribution and SYTO-61 fluorescence distribution. The time change is indicated by the mean parasite age (i.e. $\mu$). The parameter values chosen for this example are $\sigma = 0.8\ \rm h\ p.i.$, $A_r = 21\ \rm h\ p.i.$, $\lambda = 0.7\ \rm h\ p.i.^{-1}$, $F_0 = 20$ (unitless), $r_1 = 0.24\ \rm h\ p.i.^{-1}$ and $r_2 = 0.56\ \rm h\ p.i.^{-1}$. The simulation starts with 4000 rings with a mean age of 10 h p.i. and standard deviation of 0.8 h p.i.. The total number of parasites is shown in black and the number of trophozoites in green.}}
\end{figure}

In preparation for its application to the experimental data, Fig.\ 3 presents a single stochastic realisation (performed using the Gillespie algorithm \cite{Gillespie1977}) of the model. We initiated the simulation with a population of $4000$ ring-stage parasites with a mean age of 10 h p.i. (and standard deviation of 0.8 h p.i.) in the second life-cycle. We assumed that $A_r = 21\ \rm h\ p.i.$, $\lambda = 0.7\ \rm h\ p.i.^{-1}$, $F_0 = 20$ (unitless), $r_1 = 0.24\ \rm h\ p.i.^{-1}$ and $r_2 = 0.56\ \rm h\ p.i.^{-1}$. Note that we chose these parameters simply to illustrate the model's behaviour. While the simulated fluorescence histograms may look similar to the experimental data, the parameters do not necessarily represent the true values. The simulation was implemented in MATLAB (version R2014b; The MathWorks, Natick, MA) and the code is provided in the \emph{Supporting Information}. The left panels in Fig.\ 3 show the continuous ageing of the parasite population and accumulation of trophozoites within the simulated population. The right panels show how the transition to a population dominated by trophozoites (with their corresponding increased SYTO-61 staining) leads to a distinct shift in the SYTO-61 fluorescence intensity histogram. The bimodality evident in the observed data (Fig.\ 2) is also evident in the simulated data.

\begin{table}[!ht]
\caption{\small{{\bf A list of model parameters} The unit h p.i.\ is an abbreviation of hour post-infection. We assume in the model that the SYTO-61 fluorescence intensity is unitless.}}
\begin{center}
\begin{tabular}{|p{1.9cm}|p{11cm}|p{1.6cm}|}
   \hline
   {\bf Model parameter} & {\bf Description} & {\bf Unit} \\
   \hline
   $\mu$ & mean of parasite age distribution at the time of data collection & h p.i. \\
   \hline   
   $\sigma$ & standard deviation of age distribution & h p.i. \\
   \hline 
   $A_r$ & the age before which the ring-to-trophozoite transition cannot occur & h p.i. \\
   \hline
   $\lambda$ & rate of conversion from ring to trophozoite & $\rm h^{-1}$ \\
   \hline    
   $F_0$ & a parameter in the age-to-signal mapping (Eq.\ \ref{eq:2}) & -- \\
   \hline 
   $r_1$ & the rate at which the age-dependent intensity of SYTO-61 staining increases in ring stage & $\rm [h\ p.i.]^{-1}$ \\
   \hline 
   $r_2$ & the rate at which the age-dependent intensity of SYTO-61 staining increases in trophozoite stage & $\rm [h\ p.i.]^{-1}$ \\
   \hline
\end{tabular}
\end{center}
\end{table}

\subsection*{Derivation of the probability density function for the SYTO-61 fluorescence distribution}

\begin{figure}[ht!]
\centering
\includegraphics[scale=1.5]{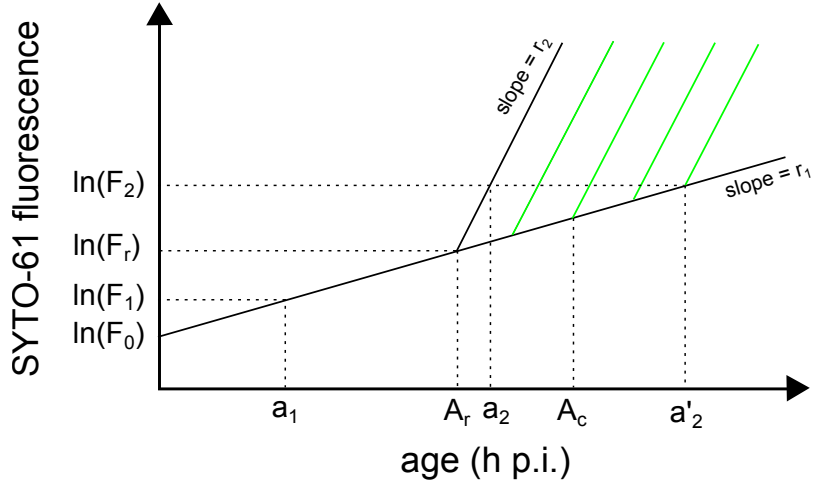}
\caption*{Figure 4: \small{Illustrative figure showing the mapping from parasite age to fluorescence intensity (in natural logarithm). If a parasite is in ring form, it will move along the line with slope $r_1$. Otherwise becoming a trophozoite at a certain age after $A_r$ will then follow a line with slope $r_2$ (e.g. the green lines).}}
\end{figure}

Our model for the transition from ring to trophozoite stage can be represented graphically by mapping parasite age to SYTO-61 fluorescence intensity (Fig.\ 4). The intensity signal for a single ring-stage parasite will increase with age along the line with slope $r_1$. On becoming a trophozoite (at a certain age $A_c > A_r$) it will then follow one of the green lines with slope $r_2$. It follows that a fluorescence intensity $F_1$ can only originate from rings with age $a_1$. However, a fluorescence intensity $F_2$ can result from trophozoites with ages between $a_2$ and $a_2^{\prime}$ (inclusive) or a ring with age $a_2^{\prime}$. These considerations suggest that we derive the probability density function for the total fluorescence signal (from a population of parasites) in a piecewise way with a critical point of separation, $F_r$, mapped from the ready-for-change age $A_r$. 

{\flushleft {\bf Case 1: $F \leq F_r$ where $F_r = F_0e^{r_1A_r}$.} Since $a_1 \sim N(\mu,\sigma^2)$ and $a_1 = {\rm ln}(F/F_0)/r_1$, we have the probability density function for $F$,
\begin{equation} 
\label{eq:3}
Pr(F) = \frac{1}{\sigma\sqrt{2\pi}}e^{-\frac{(a_1-\mu)^2}{2\sigma^2}}\left(\frac{1}{r_1F}\right),
\end{equation} 
where $1/(r_1F)$ is the derivative of $a_1$ with respect to $F$ due to use of the logarithmic transform on the fluorescence data.

{\bf Case 2: $F > F_r$ where $F_r = F_0e^{r_1A_r}$.} The probability density function for $F$ is a sum of two parts. One is the probability of rings with age $a_2^{\prime}$ and the other is the probability of trophozoites with a range of ages from $a_2$ to $a_2^{\prime}$. The former is given by 
\begin{equation} 
\label{eq:4}
Pr[F,\ {\rm from\ rings\ with\ age\ } a = a_2^{\prime}] = \frac{1}{\sigma\sqrt{2\pi}}e^{-\frac{(a_2^{\prime}-\mu)^2}{2\sigma^2}}\left[e^{-\lambda(a_2^{\prime}-A_r)}\right]\left(\frac{1}{r_1F}\right).
\end{equation} 
The first term in this expression comes from the normal distribution of the population. The second term follows from the modelled Poisson distribution, $Pois(\lambda)$, for the probability that a parasite of age $a_2^{\prime}$ remains in ring form. The last term is the derivative of $a_2^{\prime}$ with respect to $F$ due to the logarithmic transform $a_2^{\prime} = {\rm ln}(F/F_0)/r_1$.

The contribution to the probability density from trophozoites is given by
\begin{equation} 
\label{eq:5}
Pr[F,\ {\rm from\ trophozoites}] = \int_{a_2}^{a_2^{\prime}} Pr(F|a) da,
\end{equation} 
where $Pr(F|a)$ is the probability density function of $F$ from trophozoites with age $a$. To derive $Pr(F|a)$, we first give the cumulative probability
\begin{equation} 
\label{eq:6}
\int_{F}^{F+\Delta F} Pr(F|a) dF= \frac{1}{\sigma\sqrt{2\pi}}e^{-\frac{(a-\mu)^2}{2\sigma^2}}\left[e^{-\lambda(A_c(F+\Delta F)-A_r)}-e^{-\lambda(A_c(F)-A_r)}\right],
\end{equation} 
where $A_c(F) = [r_2a-{\rm ln}(F/F_0)]/(r_2-r_1)$ and $A_c(F+\Delta F) = [r_2a-{\rm ln}((F+\Delta F)/F_0)]/(r_2-r_1)$. The difference of two exponentials indicates the probability of trophozoites with conversion age between $A_c(F)$ and $A_c(F+\Delta F)$. By multiplying $1/\Delta F$ on both sides of Eq.\ \ref{eq:6} and taking the limit $\Delta F \to 0$, we obtain
\begin{equation} 
\label{eq:7}
Pr(F|a)= \frac{1}{\sigma\sqrt{2\pi}}e^{-\frac{(a-\mu)^2}{2\sigma^2}}\left[\lambda e^{-\lambda(A_c-A_r)}\right]\frac{1}{(r_2-r_1)F},
\end{equation} 
where $A_c = [r_2a-{\rm ln}(F/F_0)]/(r_2-r_1)$. We notice that the second term $\lambda e^{-\lambda(A_c-A_r)}$ is exactly the probability density function of the Poisson distribution, and the last term is the derivative of $A_c$ with respect to $F$ (except for a negative sign cancelled during simplification). 
}

Hence, the probability density function for $F$ is given by
\begin{equation}
\label{eq:8}
Pr(F) = \left\{\def\arraystretch{1.2}%
  \begin{array}{@{}c@{\quad}l@{}}
     \frac{1}{\sigma\sqrt{2\pi}}e^{-\frac{(a_1-\mu)^2}{2\sigma^2}}\left(\frac{1}{r_1F}\right), & F \leq F_r \\
     \frac{1}{\sigma\sqrt{2\pi}}e^{-\frac{(a_2^{\prime}-\mu)^2}{2\sigma^2}}\left[e^{-\lambda(a_2^{\prime}-A_r)}\right]\left(\frac{1}{r_1F}\right) +\int_{a_2}^{a_2^{\prime}} Pr(F|a) da, & F > F_r \\
  \end{array}\right.
\end{equation}
where $Pr(F|a)$ is given by Eq.\ \ref{eq:7}, $F_r = F_0e^{r_1A_r}$, $a_1 = {\rm ln}(F/F_0)/r_1$, $a_2 = [{\rm ln}(F/F_0)+(r_2-r_1)A_r]/r_2$, $a_2^{\prime} = {\rm ln}(F/F_0)/r_1$ and $A_c = [r_2a-{\rm ln}(F/F_0)]/(r_2-r_1)$. Although $a_1$ and $a_2^{\prime}$ have the same expression, we retain both for clarity. 

\subsection*{Parameter identifiability}

To apply the analytical expression for the SYTO-61 fluorescence distribution (Eq.\ \ref{eq:8}) to data, we must determine if the equation is well-posed, i.e.\ is there sufficient information within the data to uniquely identify the model's parameters. Eq.\ \ref{eq:8} contains seven parameters, but as we will now show, this equation can be reduced to an expression containing just five parameters. This expression cannot be further reduced and so is  sufficient for parameter estimation (Table 2). The relationship between these five parameters and the seven ``biological'' parameters---with which we are fundamentally concerned---is derived.

We begin with the case $F < F_r$ and identify two new parameters,
\begin{equation} 
\label{eq:9}
K_1 = {\rm ln}(F_0) + r_1\mu,
\end{equation} 
\begin{equation} 
\label{eq:10}
K_2 = \sigma r_1.
\end{equation} 
$Pr(F)$ can be rearranged to be a two-parameter distribution
\begin{equation} 
\label{eq:11}
Pr(F) = \frac{1}{(\sigma r_1)\sqrt{2\pi}}e^{-\frac{[{\rm ln}(F) - ({\rm ln}(F_0)+r_1\mu)]^2}{2(\sigma r_1)^2}}\left(\frac{1}{F}\right) = \frac{1}{K_2\sqrt{2\pi}}e^{-\frac{[{\rm ln}(F) - K_1]^2}{2K_2^2}}\left(\frac{1}{F}\right),
\end{equation} 
where $K_1$ and $K_2$ represent the mean and standard deviation respectively. This is consistent with the fact that a log-normal distribution is uniquely determined by two parameters.

Secondly, for the case of $F > F_r$, the first part is also a log-normal distribution except for a Poisson probability density function. Thus we introduce 
\begin{equation} 
\label{eq:12}
K_3 = {\rm ln}(F_0) + r_1A_r,
\end{equation} 
\begin{equation} 
\label{eq:13}
K_4 = \frac{\lambda}{r_1},
\end{equation} 
such that the first part is given by
\begin{equation} 
\label{eq:14}
\frac{1}{\sigma\sqrt{2\pi}}e^{-\frac{(a_2^{\prime}-\mu)^2}{2\sigma^2}}\left[e^{-\lambda(a_2^{\prime}-A_r)}\right]\left(\frac{1}{r_1F}\right) = \frac{1}{K_2\sqrt{2\pi}}e^{-\frac{[{\rm ln}(F)-K_1]^2}{2K_2^2}}\left[e^{-K_4({\rm ln}(F)-K_3)}\right]\left(\frac{1}{F}\right).
\end{equation} 

Thirdly, for the integral part, we introduce
\begin{equation} 
\label{eq:15}
K_5 = \frac{r_2}{r_1}.
\end{equation} 
Therefore, Eq.\ \ref{eq:7} becomes
\begin{equation} 
\label{eq:16}
Pr(F|a)= \frac{1}{K_2\sqrt{2\pi}}e^{-\frac{(r_1a+{\rm ln}(F_0)-K_1)^2}{2K_2^2}}\left(e^{-\Phi}\right)\frac{K_4r_1}{(K_5-1)F},
\end{equation} 
where 
\[ \Phi = \frac{K_4[K_5(r_1a+{\rm ln}(F_0))-{\rm ln}(F)+K_3-K_3K_5]}{K_5-1}.\]
Then taking the transform $b = r_1a+{\rm ln}(F)$, the integral part in Eq.\ \ref{eq:8} becomes
\begin{equation} 
\label{eq:17}
\int_{a_2}^{a_2^{\prime}} Pr(F|a) da = \int_{b_2}^{b_2^{\prime}} Pr(F|b) db= \int_{b_2}^{b_2^{\prime}}\frac{1}{K_2\sqrt{2\pi}}e^{-\frac{[b-K_1]^2}{2K_2^2}}\left(e^{-\Phi}\right)\frac{K_4}{(K_5-1)F}db,
\end{equation} 
where 
\[ \Phi = \frac{K_4[K_5b-{\rm ln}(F)+K_3-K_3K_5]}{K_5-1},\]
and the lower and upper bounds are given by
\[ b_2 = \frac{{\rm ln}(F)-K_3+K_3K_5}{K_5} \qquad {\rm and} \qquad b_2^{\prime} = {\rm ln}(F).\]

At last, we find $F_r  = e^{K_3}$ which finalises the process of expressing the fluorescence distribution using five parameters, each of which should be identifiable by application of the model to the fluorescence intensity histogram data. The final form of the model is given by
\begin{equation}
\label{eq:18}
Pr(F) = \left\{\def\arraystretch{1.2}%
  \begin{array}{@{}c@{\quad}l@{}}
     \frac{1}{K_2\sqrt{2\pi}}e^{-\frac{[{\rm ln}(F) - K_1]^2}{2K_2^2}}\left(\frac{1}{F}\right), & F \leq e^{K_3} \\
     \frac{1}{K_2\sqrt{2\pi}}e^{-\frac{({\rm ln}(F)-K_1)^2}{2K_2^2}}\left[e^{-K_4({\rm ln}(F)-K_3)}\right]\left(\frac{1}{F}\right) +\int_{b_2}^{b_2^{\prime}} Pr(F|b) db, & F > e^{K_3} \\
  \end{array}\right.
\end{equation}
where the integral of $Pr(F|b)$ is given by Eq.\ \ref{eq:17}. A summary of the five new parameters is provided in Table\ 2. 

\begin{table}[!ht]
\caption{\small{{\bf A list of new parameters in the probability density function of the reduced fluorescence intensity distribution.} The new parameters are some combinations of the original model parameters listed in Table 1 and are primarily used for data fitting. The choice of parameter constraints are provided in the main text.}}
\begin{center}
\begin{tabular}{|p{3cm}|p{9cm}|p{2cm}|}
    \hline
   {\bf New parameter} & {\bf Relationship to the model parameters in Table 1} & {\bf Constraint}\\
   \hline
   $K_1$ & ${\rm ln}(F_0) + r_1\mu$ & [6.2, 10.8] \\
   \hline   
   $K_2$ & $\sigma r_1$ & [0.0039, 1] \\
   \hline 
   $K_3$ & ${\rm ln}(F_0) + r_1A_r$ & [2, 10.8] \\
   \hline
   $K_4$ & $\lambda/r_1$ & [0, 300] \\
   \hline    
   $K_5$ & $r_2/r_1$ & (1, 10] \\
   \hline 
\end{tabular}
\end{center}
\end{table}

\subsection*{Method for data fitting}

The simplified 5-parameter probability density function (Eq.\ \ref{eq:18}) was fitted to the observed SYTO-61 fluorescence data for the 3D7 strain shown in Fig.\ 2 and Fig.\ S1. While drug induces growth retardation of parasites in the first life cycle (from the time of drug exposure to the end of that cycle), the second life cycle is drug-free and assumed to progress normally. Accordingly, when modelling parasites in the second life cycle, the model parameter that may depend upon drug concentration is the mean parasite age $\mu$ (and thus $K_1$ in the 5-parameter model). Moreover, since the parasite age distribution may change over time in the experiment, the spread of parasite age distribution $\sigma$ (and in turn $K_2$) is also allowed to vary. To simultaneously fit the model to all the data shown in Fig.\ 2 (or Fig.\ S1), we therefore require 35 parameters (16 for $K_1$, 16 for $K_2$, and one for each of $K_3$, $K_4$ and $K_5$). 

To optimise the data fitting procedure, we provide plausible ranges for each parameter (summarised in Table\ 1). Based on the observed fluorescence range of approximately 500--50000 in the observed data, we have $e^{K_1} \in [500, 50000]$ which gives $K_1 \in [6.2, 10.8]$. We estimate $\sigma \in [0.39, 2]$ The lower bound is based on the the estimated standard deviation of the initial parasite age distribution where approximately 80\% of parasites were aged within a one-hour window centred at the mean age. The upper bound is chosen to allow for a much wider age distribution. $r_1$ is assumed to be in $[0.01, 0.5]$ where the lower bound avoids a very small denominator in $K_4$ and $K_5$. These give $K_2 \in [0.0039, 1]$. We have $K_3 \in [2, 10.8]$ based on a similarity to $K_1$ but with a much smaller lower bound in order to capture possible small values of $A_r$. $K_4 \in [0, 300]$ corresponds to $\lambda \in [0, 3]$ where the upper bound $\lambda = 3$ represents a very fast ring-to-trophozoite transition (over 95\% of rings become trophozoites within one hour of ageing to the ready-for-change age $A_r$). We have $K_5 \in (1, 10]$ where the lower bound is based on the assumption that $r_2 > r_1$ and the upper bound is arbitrarily chosen.

A global search using a combination of Latin Hypercube Sampling (LHS) and non-linear least squares optimisation was performed. A minimum of 4000 parameter sets (each of which contains 35 parameter values, 16 identical values for $K_1$, 16 identical values for $K_2$, and one value for each of $K_3$, $K_4$ and $K_5$) were randomly generated using MATLAB's built-in LHS function \emph{lhsdesign} (plus a few additional sets proposed based on our prior knowledge) and were used as initial points for multi-start optimisation.  The precision for the initial values was up to the first decimal place for $K_2$ and integer values were sampled for all other parameters. For each of these initial points (minimum 4000) generated using the LHS sampling, MATLAB's built-in functions \emph{lsqcurvefit} and \emph{nlparci} with default settings were used to minimise the sum of squared residuals subject to the above constraints and produce the corresponding best-fit parameter values and associated 95\% confidence intervals (CIs). The globally optimal solution and associated best-fit parameter values (which are presented in the \emph{Results}) were given by the parameter set producing the smallest sum of squared residuals. The trapezoidal integration method with a division of 1000 equal sub-intervals was used to approximate the integral over $[b_2, b_2^{\prime}]$ in the probability density function.

\section*{Results}

\subsection*{Quantifying the dependence of drug-induced parasite growth retardation on drug concentration}

Fig.\ 5 presents the results of fitting the 5-parameter model to the SYTO-61 fluorescence data for ART (the best-fit parameter values are provided in Supplementary Table S1).  The model correctly captures the ART-dependent change of shape in SYTO-61 fluorescence histograms --- increasing ART concentration leads to an increase in the ring population (i.e. the mode with the lower fluorescence) at the expense of a decreased trophozoite population (i.e. the mode with the higher fluorescence). 

\begin{figure}[ht!]
\centering
\includegraphics[scale=1]{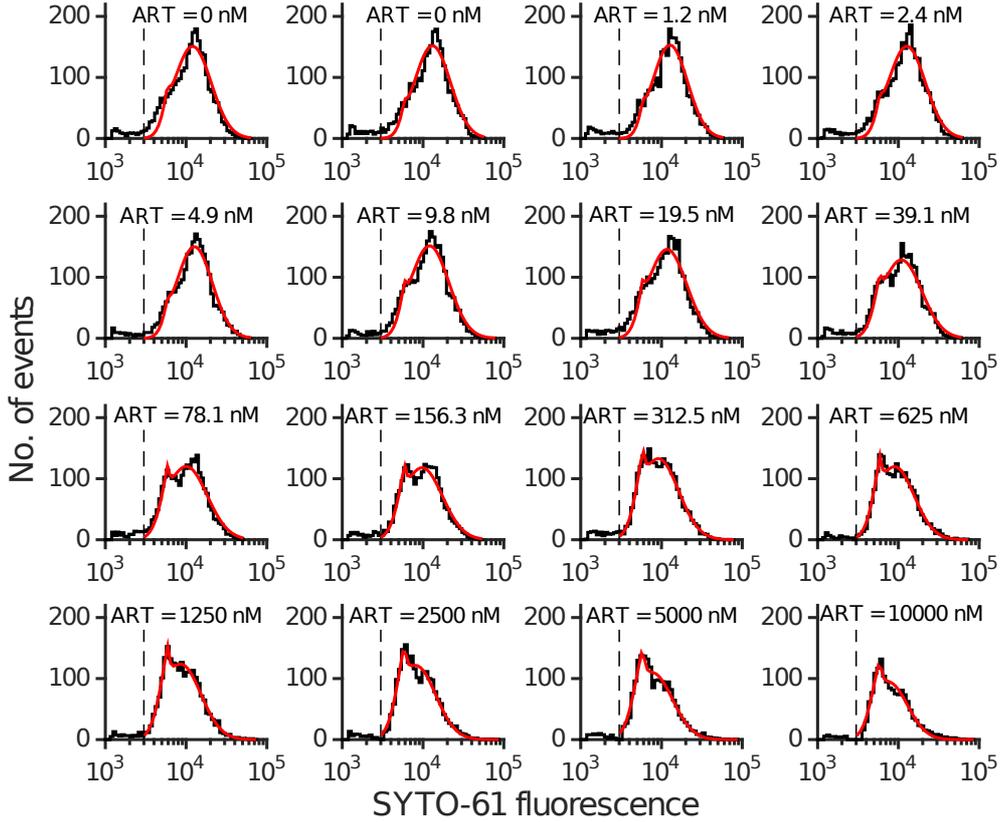}
\caption*{Figure 5: \small{Results of fitting the model to SYTO-61 fluorescence frequency data with various ART concentrations (note that two cultures with 0 nM ART were measured). The red curves represent the best-fits to the histogram data. For each panel, the samples with intensity less than 3000 (indicated by the dashed lines) were excluded from the model fitting process. Note that the histograms are corrected by removing the unviable population (see \emph{Materials and Methods} for details). Full fitting results are provided in Supplementary Table S1.}}
\end{figure}

To quantify the extent of ART-induced parasite growth retardation, we define $T_{\Delta} = \mu - A_r$, the time period (in the second life cycle) from the \emph{ready-for-change} age $A_r$ to the mean population age at the time of data collection $\mu$. Recapping Fig. 1, since the delay in rupture time of the first cycle directly affects the mean parasite age at the time of data collection $\mu$, but not the \emph{ready-for-change} age $A_r$, the change in $T_{\Delta}$ therefore provides a direct indication of how much the rupture time is delayed in the first life cycle due to drug application. A smaller $T_{\Delta}$ indicates a longer delay in the rupture time (in the first cycle), i.e.\ parasites are younger when measured. Conversely, a larger $T_{\Delta}$ indicates a shorter delay, i.e.\ parasites are relatively older at the time of data collection. Given that $r_1T_{\Delta} = K_1 -  K_3$ and that the age-dependent SYTO-61 staining rate $r_1$ is independent of drug concentration, we can use $K_1-K_3$ to quantify the dependence of growth retardation on drug concentration. A smaller value for $K_1-K_3$ indicates a longer delay in the rupture time (in the first cycle). Fig.\ 6A shows that, for an ART concentration of less than 5~nM (applied as a 4h pulse), little if any growth retardation is evident. However, when ART concentration increases above 5~nM, $K_1-K_3$ (and so $T_{\Delta}$) decreases approximately linearly, implying an approximately linear positive correlation between the delay to rupture time (for the first cycle) and drug concentration.      

\begin{figure}[ht!]
\centering
\includegraphics[scale=1.1]{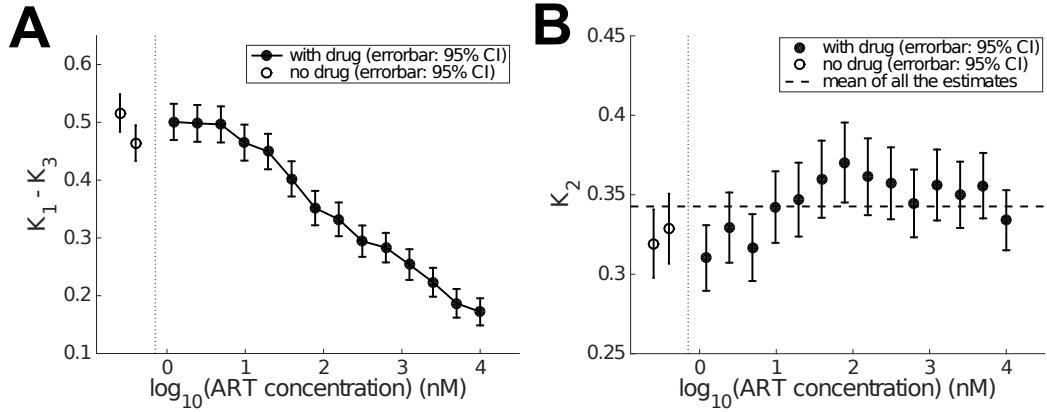}
\caption*{Figure 6: \small{Dependence of growth retardation and spread of parasite age distribution on applied ART concentration. The extent of growth retardation is indicated by $K_1-K_3$. A smaller value for $K_1-K_3$ indicates a longer delay in the rupture time in the first cycle. The extent of spread of parasite age distribution is indicated by $K_2$. A larger $K_2$ indicates a wider parasite age distribution. The horizontal dashed line indicate the mean of all the estimates of $K_2$. The vertical dotted lines separate the results of ``no drug" cases and ``with drug" cases (for a proper display of the ``no drug" cases).}}
\end{figure}

For clinically relevant DHA, the model performs similarly well (Fig.\ 7). Similar to ART, we observe a region where parasite growth is almost unaffected (e.g. DHA less than 1 nM; see Fig.\ 8A) followed by a region where $K_1-K_3$ decreases monotonically with increasing DHA concentration. However, unlike for ART, the relationship is non-linear. For high drug concentration, $K_1-K_3$ ceases to decrease (although this might be inaccurate due to limited numbers of samples caused by low viabilities), indicating some sort of saturation effect.    

\begin{figure}[ht!]
\centering
\includegraphics[scale=1]{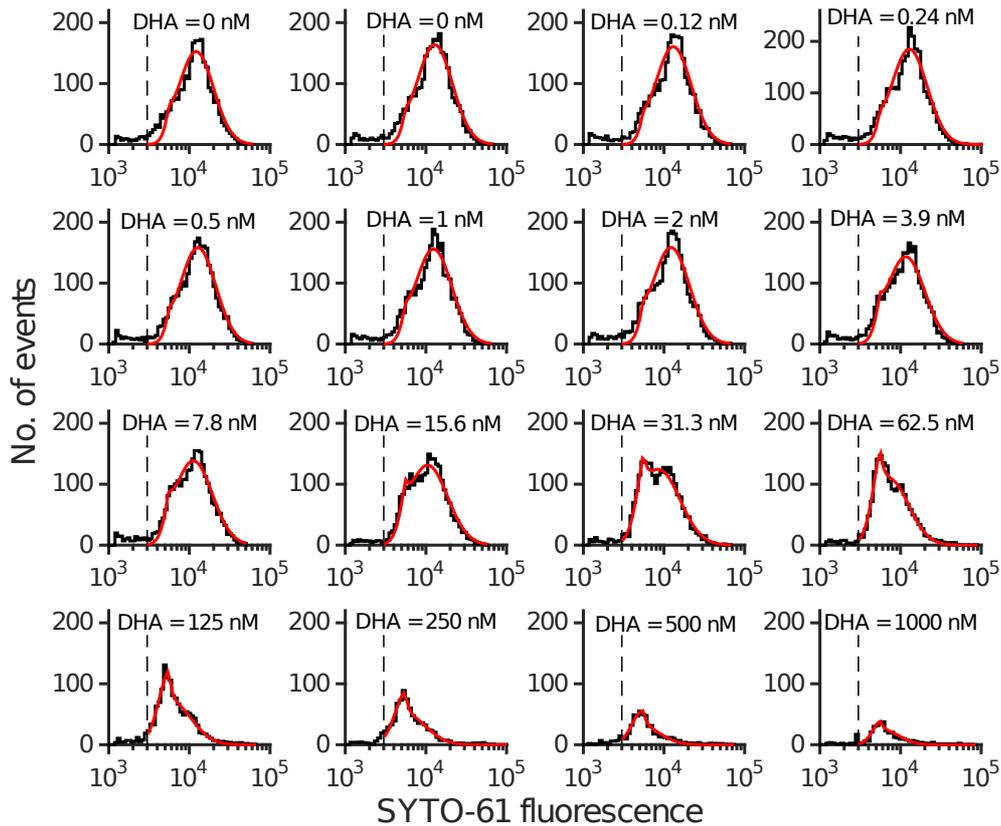}
\caption*{Figure 7: \small{Results of fitting the model to experimental data of SYTO-61 fluorescence frequency histograms with various DHA concentrations (note that two cultures with 0 nM DHA were measured). The red curves represent the best-fits to the histogram data. For each panel, the samples with intensity less than 3000 (indicated by the dashed lines) were excluded in model fitting process. Note that the histograms are corrected by removing the unviable population (see \emph{Materials and Methods} for details). Full fitting results are provided in Supplementary Table S2.}}
\end{figure}

Although we cannot determine $T_{\Delta}$ from $K_1-K_3$ due to the unknown parameter $r_1$ (the rate of age-dependent SYTO-61 staining in ring stage), we can provide some useful bounds based on plausible values for $r_1$. For example, if $r_1$ is 0.1 $\rm [h\ p.i.]^{-1}$ then an increase in ART concentration from 0 nM to approximately 1000 nM leads to a decrease of $K_1-K_3$ from approximately 0.5 to 0.26 (Fig.\ 6A), implying a delay in rupture time of approximately $T_{\Delta} = (0.5-0.26)/0.1 = 2.4$ hours. Similarly, for the same $r_1$ value but an increase in DHA concentration from 0 nM to 100 nM (see Fig.\ 8A), the estimated delay in rupture time is approximately $((0.5-0.1)/0.1 = )$ 4 hours, which is longer than that induced by 1000nM ART. Clearly, if $r_1$ is smaller (or larger), the actual rupture time will be delayed (advanced) accordingly. Based on a reasonable range for $r_1$ of [0.05--0.25] $\rm h\ p.i.^{-1}$, the drug-induced delay in rupture time (for the first life cycle) should be no more than 10 hours.    

\subsection*{The parasite age distribution is relatively unaffected by the drug pulse}

The parameter $K_2$ ($=\sigma r_1$) directly indicates how the standard deviation (or spread) of the parasite age distribution at the time of data collection ($\sigma$) depends on drug concentration. As shown in Figs.\ 6B and 8B, there is a weak positive correlation for relatively low drug concentrations. For relatively high drug concentrations, although $K_2$ drops substantially for DHA (Fig.\ 8B, which might be strongly affected by small sample size), no substantial decrease is observed for ART (Fig.\ 6B). These results suggest that, although varying drug concentration may affect the spread of the parasite age distribution, the effect is usually very limited.

\begin{figure}[ht!]
\centering
\includegraphics[scale=1.1]{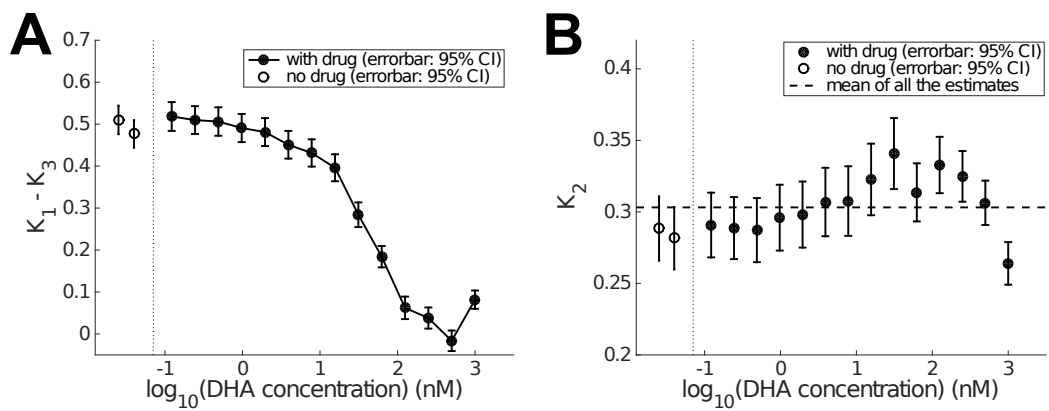}
\caption*{Figure 8: \small{Dependence of growth retardation and spread of parasite age distribution (indicated by $K_2$) on applied DHA concentration. The extent of growth retardation is indicated by $K_1-K_3$. A smaller value for $K_1-K_3$ indicates a longer delay in the rupture time in the first cycle. The extent of spread of parasite age distribution is indicated by $K_2$. A larger $K_2$ indicates a wider parasite age distribution. The horizontal dashed line indicates the mean of all the estimates of $K_2$. The vertical dotted line separates the results of ``no drug'' cases and ``with drug'' cases.}}
\end{figure}

\subsection*{New experiments to test alternative mechanisms for drug-induced growth retardation}

\begin{figure}[ht!]
\centering
\includegraphics[scale=1]{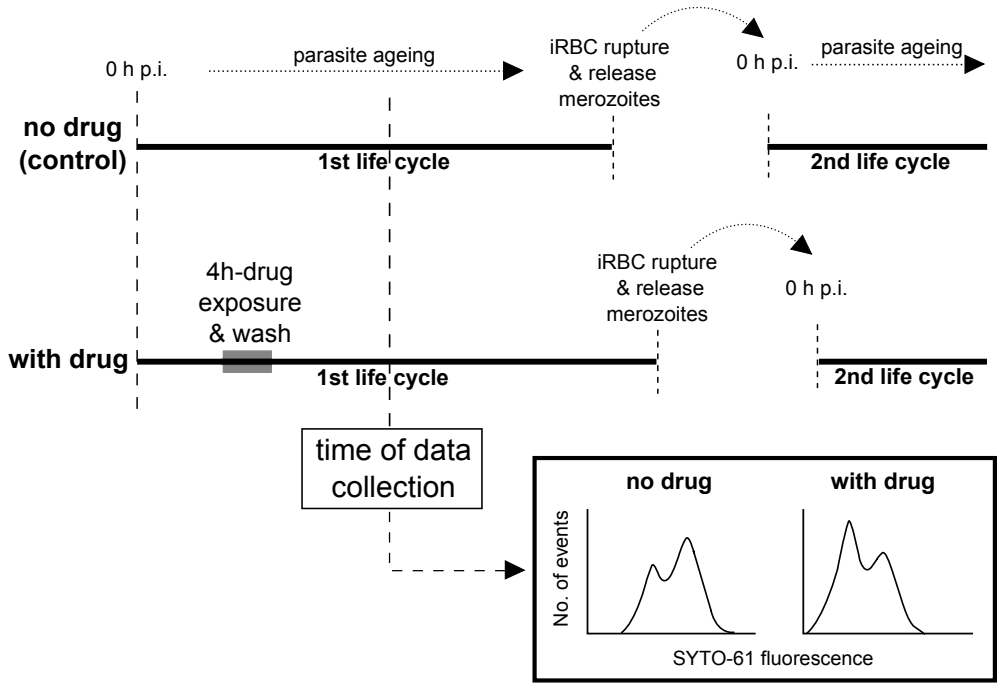}
\caption*{Figure 9: \small{Schematic diagram showing the proposed experiment for the identification of the possible mechanisms of growth retardation, i.e. a decreased transition rate $\lambda$ or a postponed \emph{ready-for-change} age $A_r$. The experiment is almost the same as the \emph{in vitro} experiment introduced in the \emph{Materials and Methods} and Fig.\ 1. The only change is that the time of data collection is moved from the first to the second life cycle, although we acknowledge potential experimental challenges due to the requirement that only the signal from viable parasites is interpreted. The exact time of data collection depends on further experimental adjustments and should capture the ring-to-trophozoite transition in the first life cycle (and not overlap with the drug application window). Measured fluorescence histogram data (again normalised) will be used for estimation of model parameters.}}
\end{figure}

Having established that our model can be used to quantify the relationship between applied drug concentration and growth retardation for the \emph{in vitro} data, we now explore the utility of the model for identifying possible alternative mechanisms responsible for growth retardation. Focusing on the ring-to-trophozoite transition (which is the core part of the model), one hypothesis is that growth retardation may be a result of either a decreased transition rate $\lambda$ or a postponed \emph{ready-for-change} age $A_r$ for parasites in the first life cycle (and thus delayed rupture time and initiation of the second life cycle). To test the hypothesis, we would require fluorescence intensity measurements for viable parasites to be performed during the first life cycle, particularly around the period of the ring-to-trophozoite transition. Such an experimental setup is illustrated in Fig.\ 9. We note that the \emph{in vitro} data used thus far, in which fluorescence measurements were taken in the second life cycle, were primarily designed to assess viability (which is, of course, only measurable through study of the second life cycle.) and so are not suitable for testing this hypothesis. We have therefore taken an \emph{in silico} approach in anticipation of future experimental studies.

\begin{figure}[ht!]
\centering
\includegraphics[scale=1]{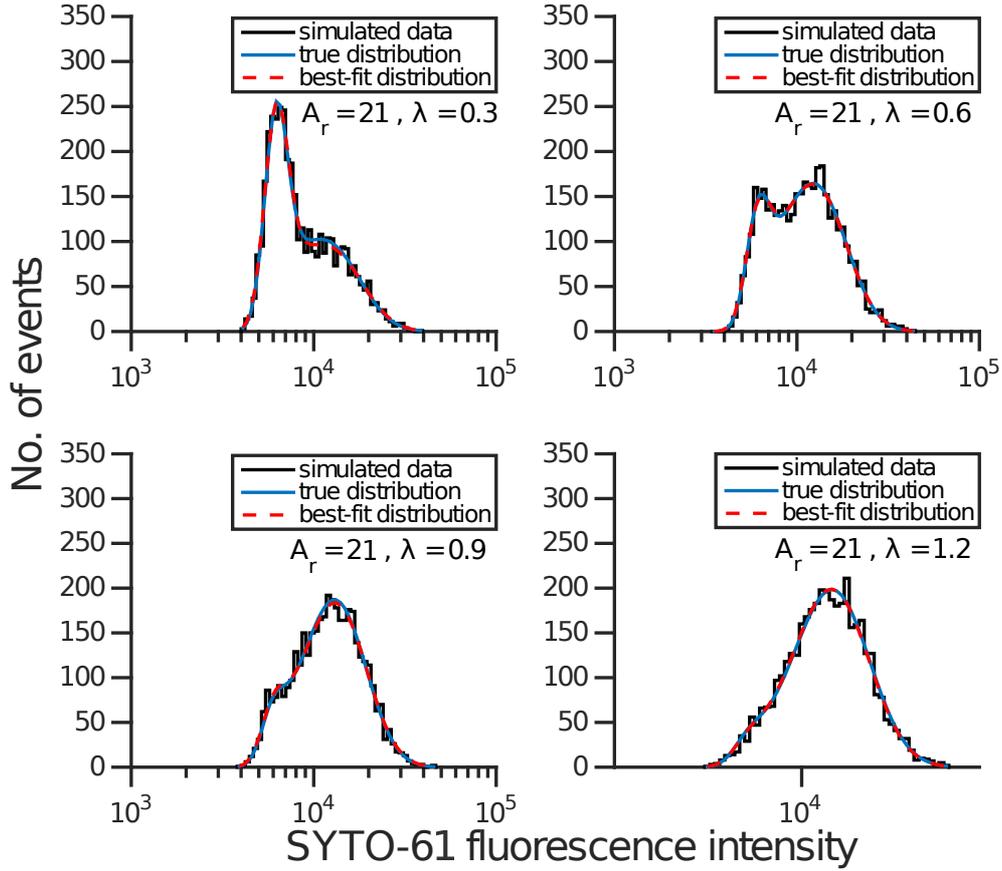}
\caption*{Figure 10: \small{Simulated data with both the true and best-fit distributions. 4000 samples were used for each histogram. The values of $A_r$ and $\lambda$ used to generate the simulated data are indicted in each panel. Other parameters are provided in Table S3. The true distribution is given by Eq.\ \ref{eq:18}. The method for data fitting is given in the \emph{Materials and Methods}.}}
\end{figure}

First, suppose that increasing ART concentration decreases the ring-to-trophozoite transition rate $\lambda$ (while leaving the \emph{ready-for-change} age $A_r$ unchanged). Fig.\ 10 shows simulated data and the model fit for four different values of the transition rate $\lambda$ between 0.3 and 1.2. Simulated data is in black (generated by assuming 4000 parasites were present) and the best-fit model in red (dashed lines). The true (model-based) distribution from which the simulated data were obtained is shown in blue. Fig.\ 11 presents estimates for $K_3$ and $K_4$, which determine $A_r$ and $\lambda$ respectively (Table 2). The model fitting procedure correctly identifies that $A_r$, the \emph{ready-for-change} age, was unchanged across the simulations, while $\lambda$, the ring-to-trophozoite transition rate, was increased. We note the presence of some minor bias in the estimate for $K_3$ for this particular simulated dataset. Supplementary Fig.\ S2 presents summary statistics for 100 replications of the simulation-reestimation procedure and establishes that unbiased estimates were obtained. The relative bias for all 11~parameters ($K_1$, $K_2$, $4 \times K_3$, $4 \times K_4$, $K_5$) was very small, the largest being 1.03\%).

\begin{figure}[ht!]
\centering
\includegraphics[scale=1.1]{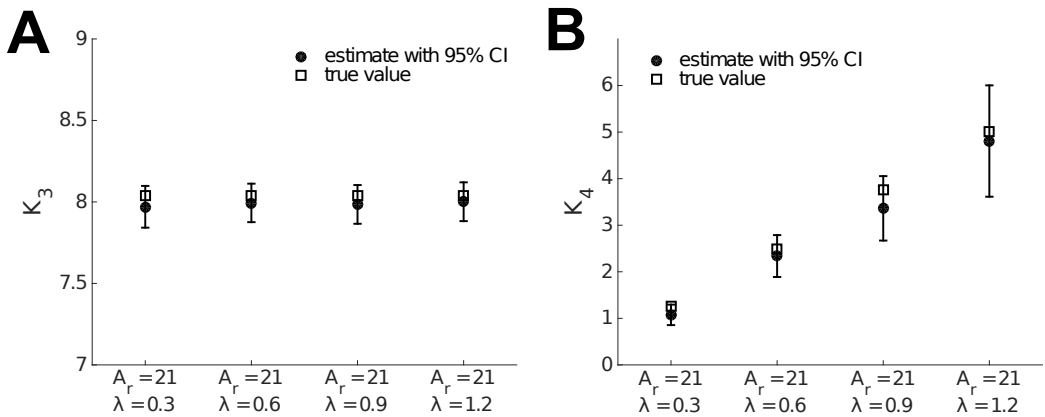}
\caption*{Figure 11: \small{Comparison of best-fit parameter values and the true values for the simulated dataset shown in Fig.\ 10. Error bars indicate 95\% CIs of the estimates. Full fitting results are given in Supplementary Table S3.}}
\end{figure}

In addition to varying $\lambda$ only, we also examined the other three logical possibilities: 1) $\lambda$ is fixed and $A_r$ is varied (see Figs.\ S3 and S4; Supplementary Table S4); 2) both $\lambda$ and $A_r$ are varied but in opposite directions, i.e.\ one increases while the other decreases (see Figs.\ S5 and S6; Supplementary Table S5); 3) both $\lambda$ and $A_r$ are varied but in the same direction, i.e. both increase or decrease (see Figs.\ S7 and S8; Supplementary Table S6). In all situations, the results consistently show that the possible variations in the key parameters $\lambda$ and $A_r$ are identified for a given set of simulated fluorescence intensity histogram data. Hence, with future availability of fluorescence data collected during the first cycle, we anticipate that our model will reliably distinguish between the alternative hypothesised mechanisms driving growth retardation.

\subsection*{Predicting the effect of growth retardation on \emph{in vivo} parasite killing}

\begin{figure}[ht!]
\centering
\includegraphics[scale=1]{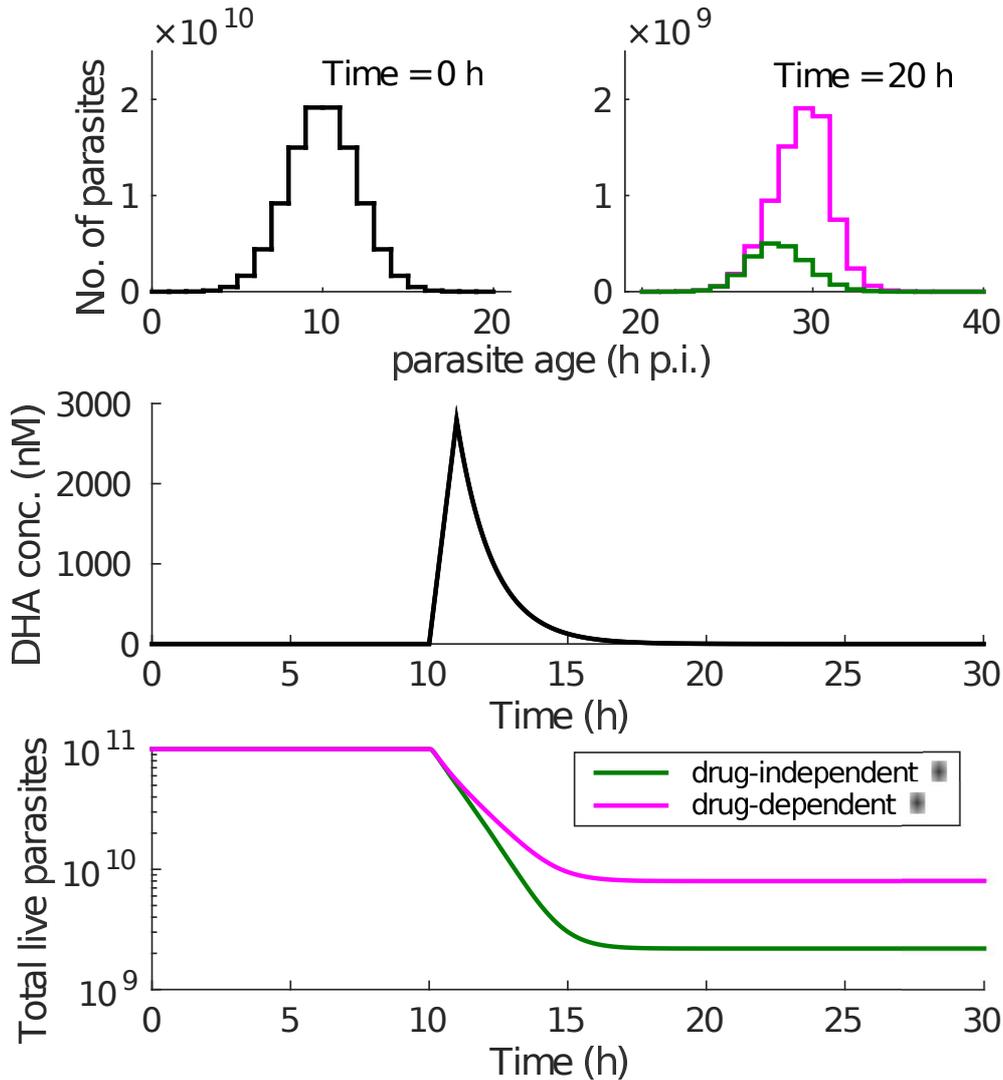}
\caption*{Figure 12: \small{Simulation of \emph{in vivo} parasite killing under a single dose of artesunate. The simulation is initiated with $10^{11}$ parasites initially distributed into 20 age bins from age 1 h p.i. to 20 h p.i. (we only consider integer ages and note that h p.i. refers to the age of iRBCs, not the time of host infection) following a normal distribution with a mean of 10 h p.i. and a standard deviation of 2 h p.i. at the start of simulation (upper left panel). We let the population grow for 10 hours' simulation time in order that a mixed population of rings and trophozoites appeared when a single dose of artesunate was applied. The pharmacokinetic profile of DHA (the active metabolite of artesunate) is given in the middle panel. If the ring-to-trophozoite transition rate $\lambda$ is assumed to be drug-independent  (i.e. there is no drug-induced growth retardation), the age distribution after 20 hours' simulation (upper right panel) and the parasite killing curve (lower panel) are shown in dark green. Results under the assumption that $\lambda$ is drug-dependent are shown in magenta.}}
\end{figure}

Although a short pulse (4 hours) of ART/DHA may prolong the parasite's life cycle by a relatively short time (e.g. less than 10 hours as estimated above), it could have a significant effect on \emph{in vivo} parasite killing. For example, if the delay occurs during the ring-to-trophozoite transition (as hypothesised above), then the recent identification that ring-stage parasites exhibit a much lower sensitivity to ART (and DHA) than those in the trophozoite stage \cite{Saralambaetal2011,Klonisetal2013,Witkowskietal2013} directly suggests that delayed ageing may lead to effective drug escape. Accordingly, here we perform a preliminary investigation of the possible impact of growth retardation on \emph{in vivo} parasite killing using a simple PK--PD model. The model contains just two compartments representing populations of rings and trophozoites, with a transition rate from rings to trophozoites given by the rate $\lambda$. Rings are killed by ART/DHA at a slower rate than trophozoites, as suggested by previous findings \cite{Saralambaetal2011,Klonisetal2013,Witkowskietal2013}. Model details are provided in the \emph{Supporting Information}.

Simulation results are shown in Fig.\ 12. A population of ring-stage parasites were initially distributed with a mean age of 10 h p.i. (note, this is simply the time since the simulated population of RBCs were infected, not the time of clinical exposure for the simulated host) and a standard deviation of 2 h at the start of simulation. With a single dose of artesunate (2mg/kg) applied at 10 hours (see the middle panel of Fig.\ 12), the rate of parasite death for a scenario in which there is no drug-induced growth retardation (i.e. $\lambda$ is independent of drug concentration) is significantly higher than that for a scenario in which there is drug-induced growth retardation (i.e. $\lambda$ is a function of drug concentration). Growth retardation delays the transition from rings to trophozoites and in turn prevents those parasites from being killed as the drug's short half-life (approximately 0.9 h) allows them to avoid exposure during the trophozoite stage. This result suggests that, if drug induces a slower transition from ring to trophozoite stage, growth retardation has a substantial adverse effect on efficient parasite killing and may thus be considered as a potential mechanism for decreasing ART sensitivity. 

\section*{Discussion}

In this paper, we have studied ART-induced parasite growth retardation using a mechanistic model that considers the ring-to-trophozoite transition to be a two-stage process and exploits the differing rates of nucleic acid production (measured through SYTO-61 fluorescence intensity) in those two life stages. By fitting the model to fluorescence histogram data, we have been able to identify the dependence of growth retardation on applied drug concentration (Figs.\ 6A and 8A) and obtain reliable estimates for how much the parasite's life span is prolonged due to exposure to drug. Our primary findings from this analysis were that: 1) drug-induced parasite growth retardation exhibits a threshold-like behaviour such that growth retardation is evident only when drug concentration is sufficiently large (for example, $>10\ \rm nM$ ART for a 4 hour drug pulse); 2) the parasite life cycle is prolonged by no more than 10 hours due to application of ART/DHA; and 3) the parasite age distribution at the time of data collection in the second life cycle is relatively unaffected by the application of short drug pulses with different concentrations.

Furthermore, we used the model to propose a hypothetical mechanism of ART-induced parasite growth retardation. We considered growth retardation to be due to either a decreased transition rate $\lambda$ or a postponed \emph{ready-for-change} age $A_r$, or a combination of the two. We have shown that, if fluorescence intensity data were collected in the first life cycle, that the model is able to accurately identify the dependencies of the transition rate $\lambda$ and the \emph{ready-for-change} age $A_r$ on drug concentration. With the availability of new experimental data (e.g.\ fluorescence data from the first life cycle), we will be able to incorporate growth retardation into our recently developed dynamic-stress model of antimalarial action \cite{Caoetal2016}, providing a comprehensive platform for the study and optimisation of ART-based therapies.

ART-induced growth retardation has two competing effects. Firstly, it slows the overall rate of growth by extending the life cycle. However, by delaying the transition to the trophozoite stage, the parasite avoids being exposed to drug at a highly sensitive stage \cite{Klonisetal2013,Caoetal2016}, providing a net benefit (to the parasite) due to the delay (as shown in Fig.\ 12). Our preliminary \emph{in vivo} simulations suggest that the overall effect is likely to be strongly beneficial (for the parasite): a short delay in the time of ring-to-trophozoite transition will have a minimal impact on the overall growth rate (and so any implications for onwards transmission are likely minimal), yet is extremely effective in avoiding the short ART/DHA pulse and so killing. Therefore, understanding where in the parasite life cycle drug-induced growth retardation acts is important in the context of combating emergent drug resistance through optimisation of dosing regimens, as demonstrated by our \emph{in vivo} simulations (Fig.\ 12). Furthermore, these results suggest that developing new longer-lived ART derivatives may overcome resistance to ART and DHA, as a direct consequence of their longer half lives.

Returning to the model we have introduced to analyse the SYTO-61 fluorescence data, it provides a major advance on current methods used to study mixed histograms from fluorescence assays. Current practice is to simply partition the bimodal signal into ``low'' and ``high'' components at a chosen threshold intensity. While suitable for well separated peaks, this method is inadequate for determining mixtures of rings and trophozoites as shown in Fig.\ 2. Our model clearly dissects the contributions of different parasite populations to the fluorescence distribution (Eq.\ \ref{eq:18}) where the integral part represents the contribution from trophozoites while the sum of all other parts represents the contribution from rings. It can therefore be used to estimate the fraction of parasite subpopulations based on   fluorescence intensity distribution in a far more reliable and rigorous way. Moreover, given that fluorescence dye staining is a standard assay in experimental biology, we anticipate that our novel methods will be broadly applicable to other problems, for example in the application to the adoptive transfer data utilised in \cite{Khouryetal2016}.

%
%
%
%
%
%
%

\section*{Funding information}
The work was supported by the National Health and Medical Research Centre of Australia (NHMRC) through Project Grants 1100394 and 1060357, the Centre for Research Excellence ViCBiostat (1035261) and the Centre for Research Excellence PRISM$^2$ (1078068). James M. McCaw was supported by an Australian Research Council (ARC) Future Fellowship. Julie A. Simpson was supported by a NHMRC Senior Research Fellowship. Leann Tilley was supported by an ARC Professorial Fellowship.

\section*{Competing interests}

We have no competing interests.

\newpage
\section*{Supporting Information}

The supporting information contains the following:
\begin{enumerate}
\item MATLAB code for the stochastic simulation shown in Fig.\ 3 in the main text
\item The model used to simulate \emph{in vivo} parasite clearance
\item MATLAB code for solving the model of \emph{in vivo} parasite clearance
\item Supplementary figures (S1--S8)
\item Supplementary tables (S1--S6)
\end{enumerate}

\newpage
\subsection*{MATLAB code for the stochastic simulation shown in Fig.\ 3 in the main text}

\begin{lstlisting}[frame=single]
clear
clc
tic

Ar = 21; % ready-for-change age
total_Num = 4e+3; % total number of viable paraistes

init_temp = 0.8*randn(1,total_Num)+10;
init_age = init_temp(init_temp>=0); % initial age distribution
age  = sort(init_age);
stage = (age>=Ar)+1; 
% stage=1: rings before age Ar; stage=2: rings after Ar; stage=3: troph
index = 1:length(age);

rand1 = rand(1,length(stage)).*(stage == 2);
Xi = log(1./rand1);
tracking = zeros(1,length(rand1)); % tracking variable

lambda0 = 0.7; % ring-to-troph transition rate
lambda = (stage == 2).*lambda0;
age_to_change = tracking; % the actual age when a ring change to troph

r1 = 0.24;
r2 = 0.56;  
F0 = 20;

SYTO61 = F0*exp(r1*age);

age0=age;
stage0=stage;
SYTO610=SYTO61;

dt = 0.01;

t = 0:dt:20;

time_to_record = 11:15; % record the results of mean age=21,22,23,24,25

indext=1:length(time_to_record);

age_all=ones(length(time_to_record),1)*age;
stage_all=ones(length(time_to_record),1)*stage;
SYTO61_all=ones(length(time_to_record),1)*SYTO61;

wb=waitbar(0,'please wait');

for i = 2:length(t)
    
    % update age
    age = age+dt;
    
    % update the status for those whose ring-to-troph transition occurs  
    tracking = tracking + lambda*dt;
    trs = (tracking-Xi>=0); % transition to troph occurs
    stage = stage+trs;
    age_to_change = age_to_change+trs.*age;
    tracking = tracking.*(1-trs); % update tracking variable
    rand1 = rand1.*(1-trs);
    
    % update the status of the rest parasites 
    temp = age.*(stage == 1);
    indtemp=index(temp>=Ar);
    rand1(indtemp) = rand(1,length(indtemp));
    Xi = log(1./rand1);
    stage(indtemp) = 2;
    lambda = (stage == 2).*lambda0;
    
    % SYTO61 fluo cumulation
    SYTO61 = (stage<3).*(F0*exp(r1*age))+...
        (stage==3).*(F0*exp(r1*age_to_change).*exp(r2*(age-age_to_change)));
    
    % recording results
    indt=indext(time_to_record==t(i));
    if ~isempty(indt)
        age_all(indt,:) = age;
        stage_all(indt,:) = stage;
        SYTO61_all(indt,:) = SYTO61;
    end
    
    waitbar(i/length(t))
end

close(wb)

toc

\end{lstlisting}
After running the above code, the following code is used to generate the result of Fig.\ 3 in the main text.
\begin{lstlisting}[frame=single]
for kk=1:5;
    
    bin_edge = 0:0.2:48;
    
    ydata_age=histcounts(age_all(kk,:),bin_edge);
    temp = age_all(kk,:);
    temp1 = stage_all(kk,:);
    troph_age = temp(temp1==3);
    ydata_trogh_age=histcounts(troph_age,bin_edge);
    
    
    % for plot only
    xnodes = zeros(1,2*length(bin_edge));
    xnodes(1:2:end) = bin_edge;
    xnodes(2:2:end) = bin_edge;
    
    ynodes = zeros(1,2*length(bin_edge));
    ynodes(2:2:end-1) = ydata_age;
    ynodes(3:2:end-1) = ydata_age;
    
    ynodes1 = zeros(1,2*length(bin_edge));
    ynodes1(2:2:end-1) = ydata_trogh_age;
    ynodes1(3:2:end-1) = ydata_trogh_age;
    
    figure(1)
    subplot(5,2,1+2*(kk-1))
    plot(xnodes,ynodes1,'-','linewidth',2,'color',[0 1 0])
    hold on
    plot(xnodes,ynodes,'-','linewidth',2,'color',[0 0 0])
    
    % xlabel('parasite age (h p.i.)')
    % ylabel('No. of events')
    box off
    
    set(gca,'xlim',[16 30],'ylim',[0 500],'fontsize',14)
    set(gca,'TickDir','out','LineWidth',2)
    set(gca,'ticklength',[0.02 0.01])
    text(20,200,['mean parasite age = ',num2str(time_to_record(kk)+10),' h p.i.'],'fontsize',14)
    
    
    % SYTO fluo histogram
    
    bin_edge = exp(linspace(log(min(SYTO61_all(kk,:))),log(max(SYTO61_all(kk,:))),50));
    
    ydata=histcounts(SYTO61_all(kk,:),bin_edge);
    
    temp = SYTO61_all(kk,:);
    temp1 = stage_all(kk,:);
    troph_fluo = temp(temp1==3);
    ydata_trogh_fluo=histcounts(troph_fluo,bin_edge);
    
    % for plot only
    xnodes = zeros(1,2*length(bin_edge));
    xnodes(1:2:end) = bin_edge;
    xnodes(2:2:end) = bin_edge;
    
    ynodes = zeros(1,2*length(bin_edge));
    ynodes(2:2:end-1) = ydata;
    ynodes(3:2:end-1) = ydata;
    
    ynodes1 = zeros(1,2*length(bin_edge));
    ynodes1(2:2:end-1) = ydata_trogh_fluo;
    ynodes1(3:2:end-1) = ydata_trogh_fluo;
    
    subplot(5,2,2*kk)
    semilogx(xnodes,ynodes1,'-','linewidth',2,'color',[0 1 0])
    hold on
    semilogx(xnodes,ynodes,'-','linewidth',2,'color',[0 0 0])
    
    % xlabel('SYTO-61 fluorescence')
    box off
    
    set(gca,'xlim',[1000 100000],'ylim',[0 400],'fontsize',14)
    set(gca,'TickDir','out','LineWidth',2)
    set(gca,'ticklength',[0.02 0.01])
end
\end{lstlisting}

\newpage
\subsection*{The model used to simulate \emph{in vivo} parasite clearance}

Here we propose a simple two-compartment model to capture both parasite killing and transition from rings to trophozoites. For a certain age $a$, we denote the number of live rings by $R$ and the number of live trophozoites by $T$. Then the equation governing the dynamics of $R$, $T$ and $a$ are given by
\begin{align}
\frac{dR}{dt} & = -k_rR-\lambda R, \tag{S1}\\
\frac{dT}{dt} & = \lambda R-k_t T, \tag{S2}\\
\frac{da}{dt} & = 1, \tag{S3}
\end{align}
where the term $\lambda R$ represents the rate of conversion from rings to trophozoites at the population level. Since we assume the transition can only occur after age $A_r$ ($A_r = 21\ \rm h\ p.i.$ is assumed in the simulation), $\lambda$ takes
\begin{equation}
\lambda = \left\{\def\arraystretch{1.2}%
  \begin{array}{@{}c@{\quad}l@{}}
    0, & a < A_r \\
    \frac{22}{34+C}, & a \geq A_r \\
  \end{array}\right.  \tag{S4}
\end{equation}
where $C$ is DHA concentration. The hyperbolic relationship is used to model a hypothetical property that a higher drug concentration would lead to a stronger delay in parasite growth. $k_r$ and $k_{t}$ are drug-induced parasite killing rates for rings and trophozoites respectively. In order to capture that killing rate for rings is significantly smaller than that for trophozoites \cite{Saralambaetal2011,Klonisetal2013,Witkowskietal2013}, we choose
\begin{align}
k_r & = \frac{0.5C^2}{C^2+200^2}, \tag{S5}\\
k_t & = \frac{2C^2}{C^2+200^2}. \tag{S6}
\end{align}
To model the pharmacokinetic profile where plasma DHA concentration usually follows a biphasic behaviour \cite{Saralambaetal2011}, we use
\begin{equation}
\frac{dC}{dt} = \left\{\def\arraystretch{1.2}%
  \begin{array}{@{}c@{\quad}l@{}}
    \frac{C_{max}}{t_m}, & t < t_m \\
    -\frac{ln(2)}{t_{1/2}} C, & t \geq t_m\\
  \end{array}\right.  \tag{S7}
\end{equation}
where $C_{max}$ is the maximum achievable concentration and $t_m$ indicates the time (since drug application) when the maximum concentration is achieved. The \emph{in vivo} DHA half-life $t_{1/2} = 0.9\ \rm h$ \cite{Yangetal2016}. To simulate the case of a single dose of artesunate (2mg/kg), we assume $C_{max} = 2820\ \rm nM$ and $t_m = 1\ \rm h$ \cite{Yangetal2016,Dondorpetal2009}.

To assess the effect of growth retardation on parasite clearance, we consider two scenarios, drug-dependent $\lambda$ and drug-independent $\lambda$. The former is modelled by using Eq.\ S4 while the latter is modelled by fixing $C$ to be zero in Eq.\ S4 (i.e. assuming constant $\lambda = 22/34 = 0.647\ \rm h^{-1}$ for $a \geq A_r$).

To simulate the model, we initially distribute $10^{11}$ parasites into 20 age bins from age 1 h p.i. to 20 h p.i. (we only consider integer ages) following a normal distribution with a mean of 10 h p.i. and a standard deviation of 2 h p.i. (see Fig.\ 12 in the main text). For each age group, we use MATLAB's bulit-in solver \emph{ode15s} to solve the model and obtain the time series of $T+R$. Then the time series of total number of live parasites is given by the sum of all 20 time series (result is shown in Fig.\ 12 in the main text). Moreover, we can also obtain the age distribution at any time (an example of $t = 20 \rm\ h$ is given in Fig.\ 12 in the main text). MATLAB code is provided in the next section.

\newpage
\subsection*{MATLAB code for solving the model of \emph{in vivo} parasite clearance}

\begin{lstlisting}[frame=single]
clear
tic

drug_app_time = 10; % time of drug application

th=0.9; % in vivo DHA half-life

init_age=10+2*randn(1e+11,1); % initial ages of the 10^11 parasites
age_dis = histcounts(init_age,0:1:20); % initial age distribution
age = 1:20; % age bins

t = 0:0.01:30;

% calculate the time series of DHA concentration
t1 = t(t<drug_app_time); % time before drug application
t3 = t(t>=drug_app_time+1); % decreasing phase of PK profile
t2 = intersect(setdiff(t,t1),setdiff(t,t3)); % increasing phase of PK profile
C1 = 0*t1;
C2 = 2820/1*(t2-drug_app_time);
C3 = C2(end)*exp(-log(2)/th*(t3-t2(end)));
C = [C1,C2,C3]; % full drug concentration profile

Numt = ones(length(age),length(t)); % matrix for total number of parasites
Numt(:,1) = age_dis'; % assign the initial age distribution

wb=waitbar(0,'please wait...');

for kk=1:length(age)
    init = [age_dis(kk);0;age(kk);0];
    for i=2:length(t)
        [~,Sol] = ode15s(@invivo_sim,[0 0.01],init);
        init = [Sol(end,1:3)';C(i-1)];
        Numt(kk,i) = Sol(end,1)+Sol(end,2);
    end
    waitbar(kk/length(age))
end

close(wb)
toc
clear init_age % remove large vector to release memory
\end{lstlisting}
where the function \emph{$invivo\_sim$} appearing in the command of \emph{ode15s} is given by

\begin{lstlisting}[frame=single]
function dy = invivo_sim(~,y)

% y = [ring;troph;age;DHA concentration]

kr=0.5*y(4)^2/(y(4)^2+200^2);
kt=2*y(4)^2/(y(4)^2+200^2);

alt=1; % drug-dependent ring-to-troph transition rate
% set alt=0 for a drug-independent ring-to-troph transition rate

if y(3)>=21
    lambda = 22/(34+alt*y(4));
else
    lambda = 0;
end

dy = zeros(4,1);

dy(1) = -kr*y(1)-lambda*y(1);
dy(2) = lambda*y(1)-kt*y(2); 
dy(3) = 1;
dy(4) = 0;
\end{lstlisting}

\newpage
\subsection*{Supplementary figures}

\begin{figure}[ht!]
\centering
\includegraphics[scale=0.67]{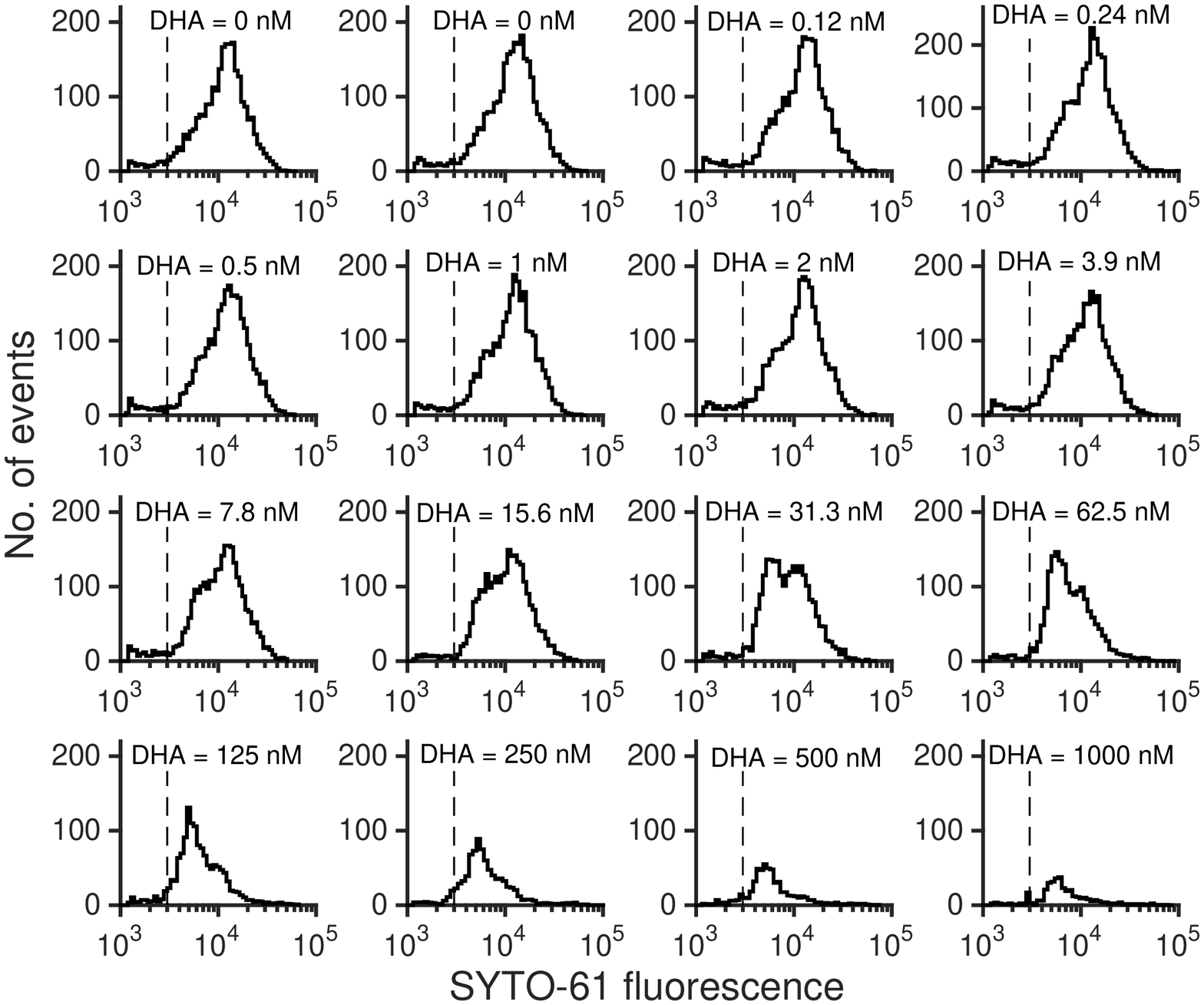}
\caption*{Figure S1: \small{Experimental data of SYTO-61 fluorescence frequency histograms with various DHA concentrations (note that two cultures with 0 nM DHA were measured). For each panel, samples with fluorescence less than 3000 (indicated by the dashed lines) were considered to include fluorescence signals from uninfected RBC and were thus not included in model fitting process. Note that the histograms are corrected by removing the unviable population (see \emph{Materials and Methods} for details).}}
\end{figure}
\newpage
\begin{figure}[ht!]
\centering
\subfloat[]{\includegraphics[scale=0.41]{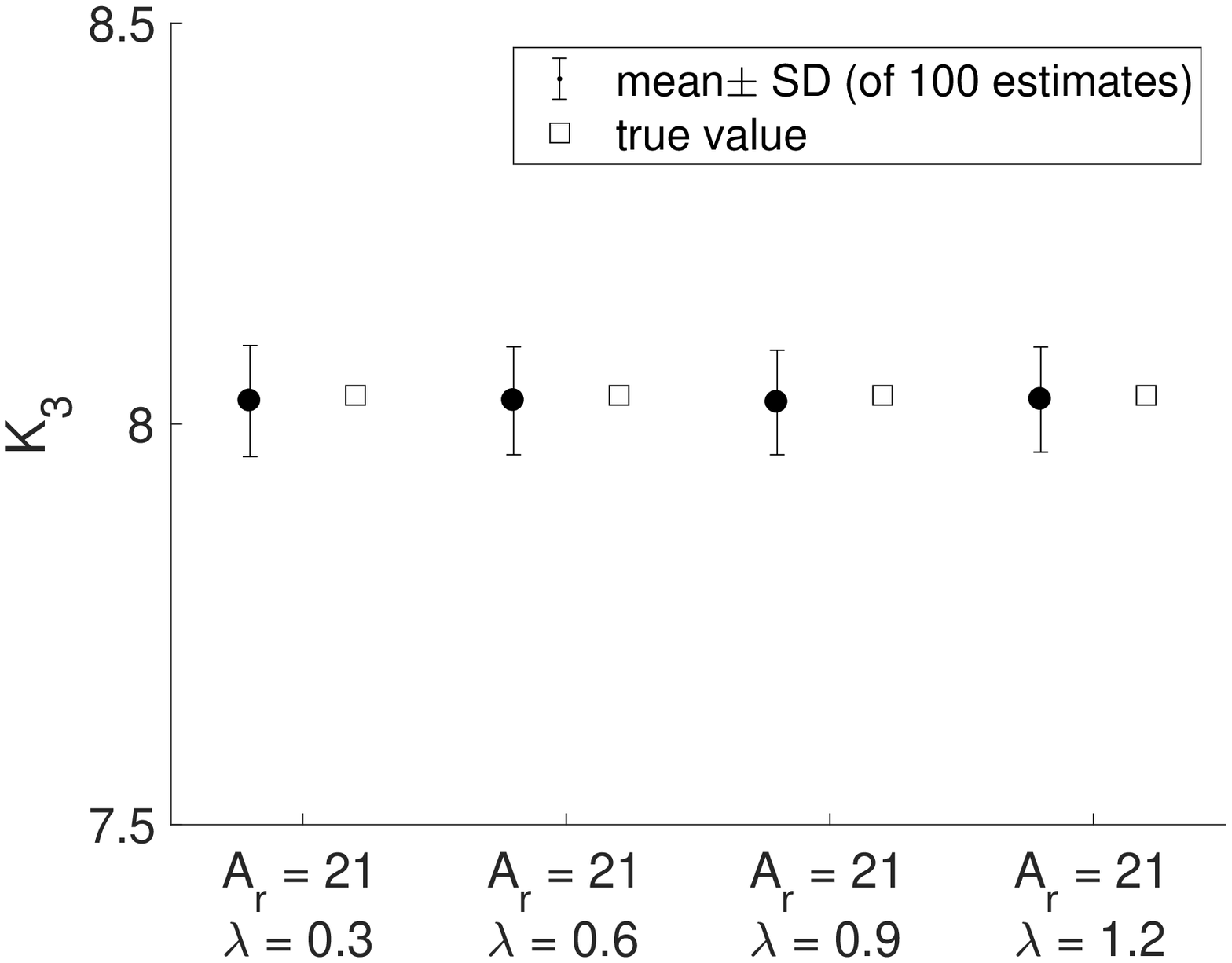}}
\subfloat[]{\includegraphics[scale=0.41]{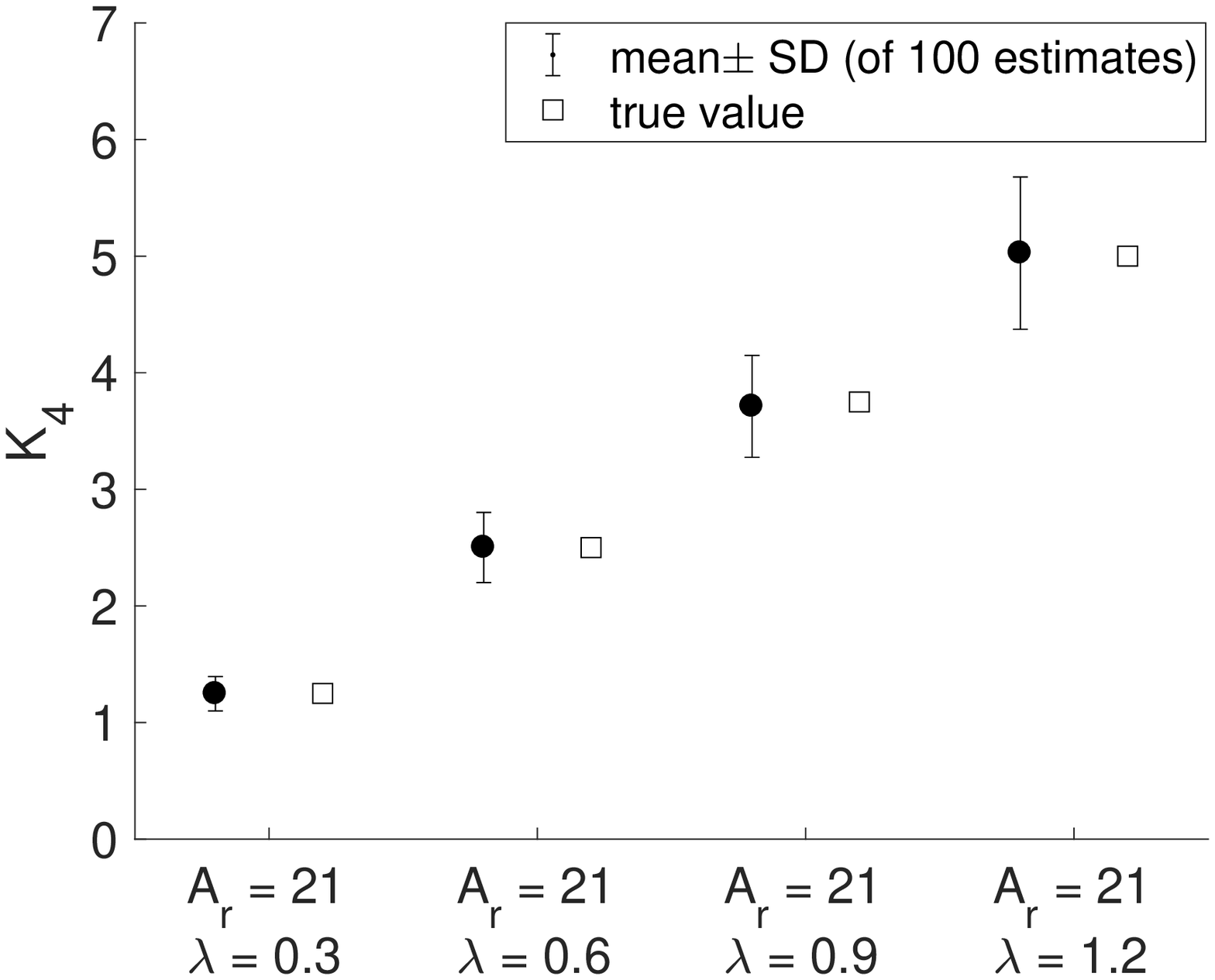}}\\
\caption*{Figure S2: \small{No systematic bias is evidenced for the simulated histogram data-based parameter estimation. In Fig. 11 in the main text, a weak underestimation of $K_3$ and $K_4$ is observed. To examine whether it is due to error in simulated histogram data (4000 samples for each histogram) or bias in the method of parameter estimation, we performed 100 trials, for each of which a set of simulated histogram data similar to Fig. 10 (generated by using the same model parameters given in Table S3) and the parameters $K_1$--$K_5$ were estimated using the same method. The figures show both the true parameter values (which are the same for the 100 trials) and the mean and SD of the estimates of the 100 trials for each parameter. Note that the error bars here represent $\rm mean\pm SD$ of the 100 estimates from the 100 trials rather than the an single estimate and 95\% CI presented in Fig. 11 in the main text. The results show that the estimates are very consistent with the true values, suggesting nearly no systematic bias (relative bias $< 1.03\%$ in this test) is induced by the fitting method.}} 
\end{figure}
\newpage
\begin{figure}[ht!]
\centering
\includegraphics[scale=0.7]{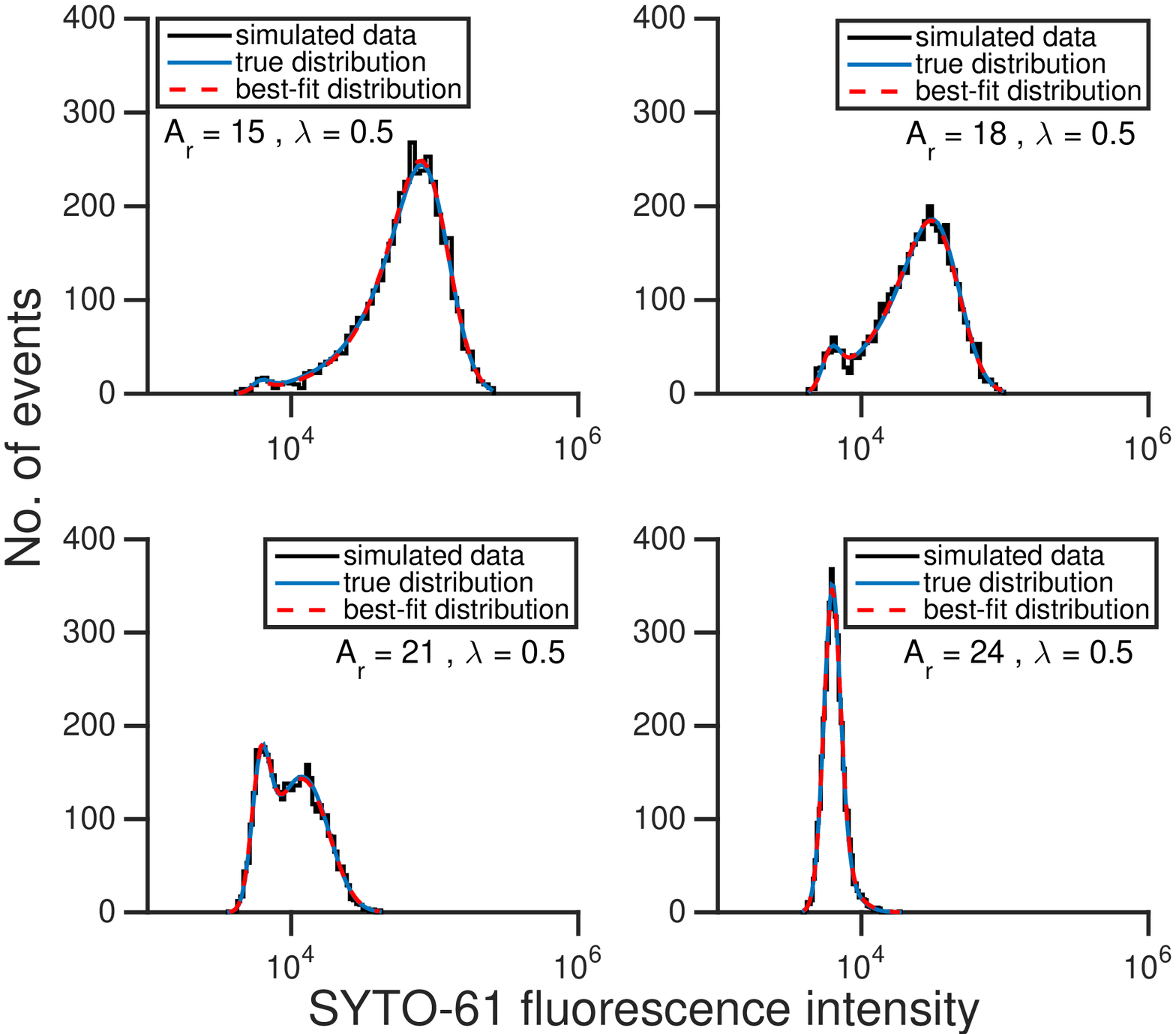}
\caption*{Figure S3: \small{Simulated data with both true distribution and best-fit distribution. 4000 samples were used for each of histograms, approximating the sample size of experimental data. The values of $A_r$ and $\lambda$ used to generate the simulated data are indicted in each panel and the other parameters are provided in Table S4.}}
\end{figure}
\newpage
\begin{figure}[ht!]
\centering
\subfloat[]{\includegraphics[scale=0.41]{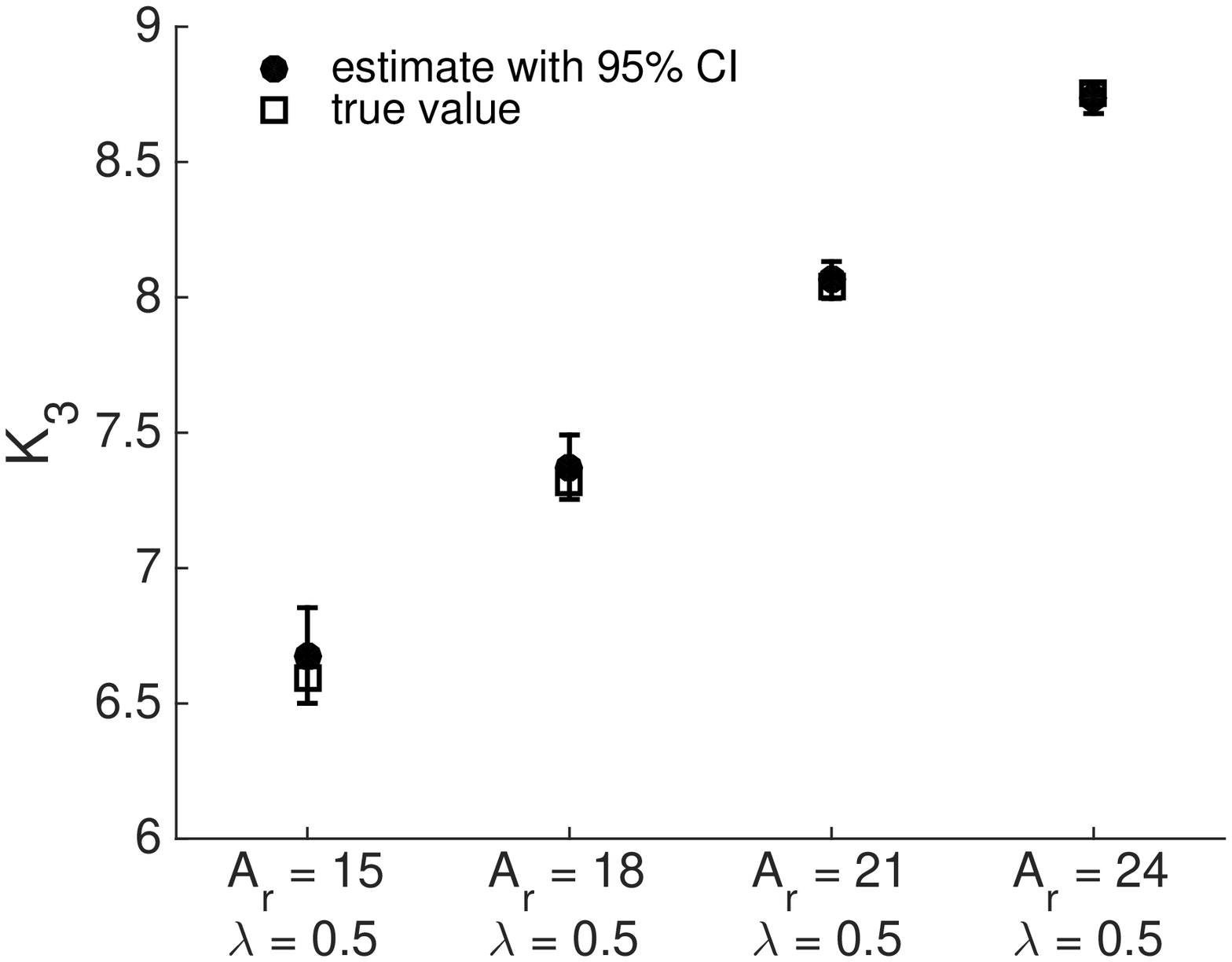}}
\subfloat[]{\includegraphics[scale=0.41]{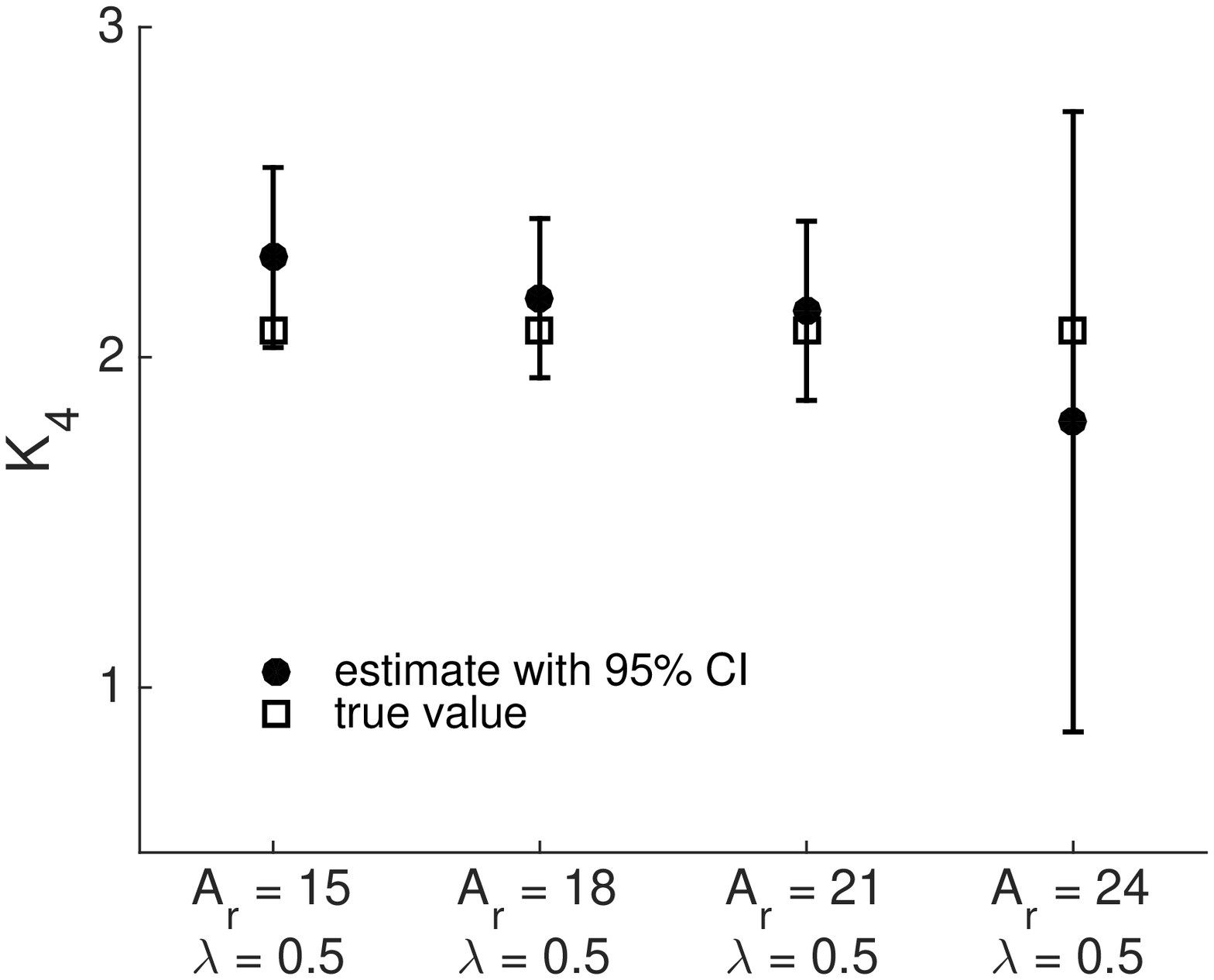}}\\
\caption*{Figure S4: \small{Comparison of best-fit parameter values and the true values for simulated data shown in Fig.\ S3. Error bars indicate 95\% CIs of the estimates. Full fitting results are given in Supplementary Table S4.}}
\end{figure}
\newpage
\begin{figure}[ht!]
\centering
\includegraphics[scale=0.7]{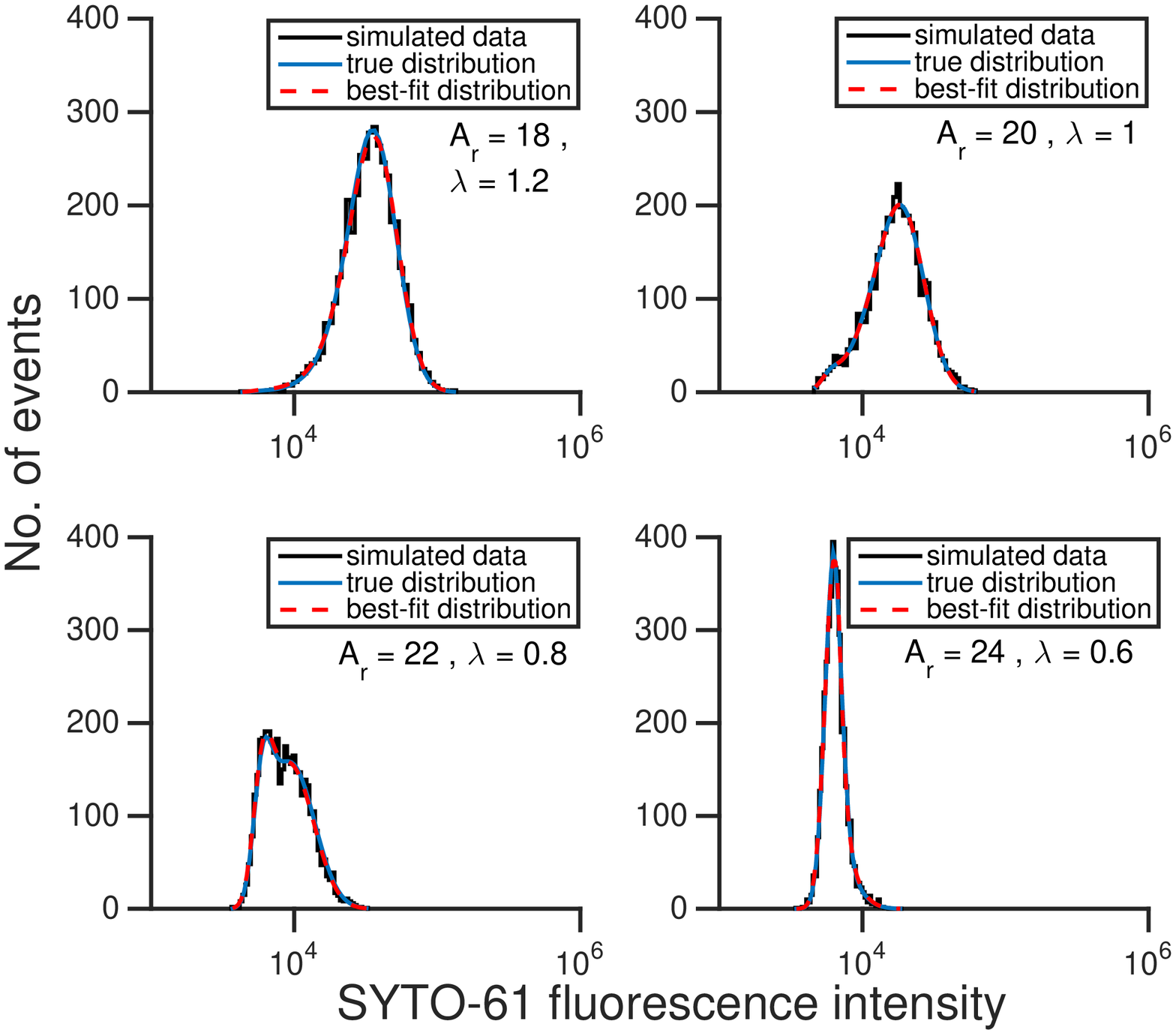}
\caption*{Figure S5: \small{Simulated data with both true distribution and best-fit distribution. 4000 samples were used for each of histograms, approximating the sample size of experimental data. The values of $A_r$ and $\lambda$ used to generate the simulated data are indicted in each panel and the other parameters are provided in Table S5.}}
\end{figure}
\newpage

\begin{figure}[ht!]
\centering
\subfloat[]{\includegraphics[scale=0.41]{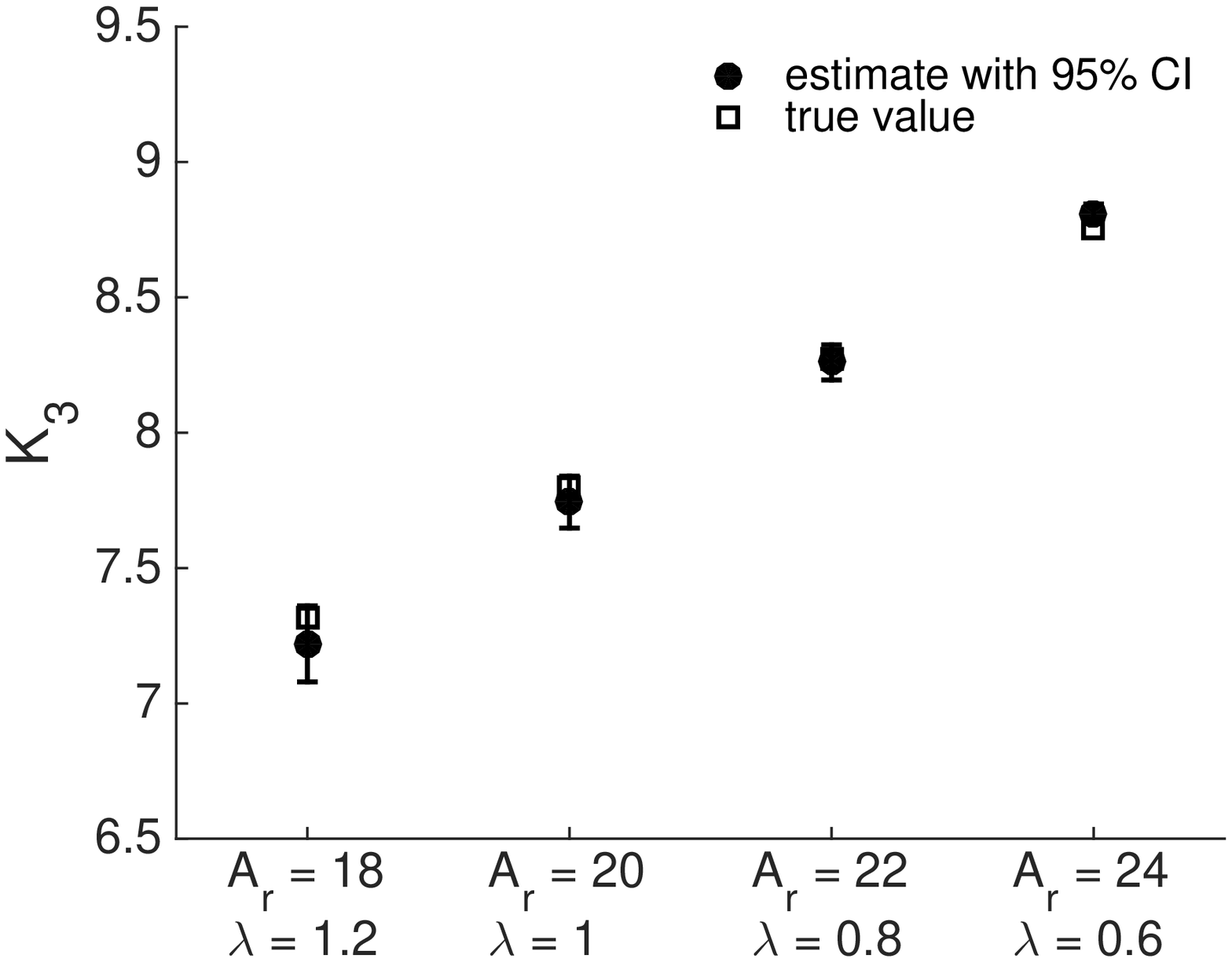}}
\subfloat[]{\includegraphics[scale=0.41]{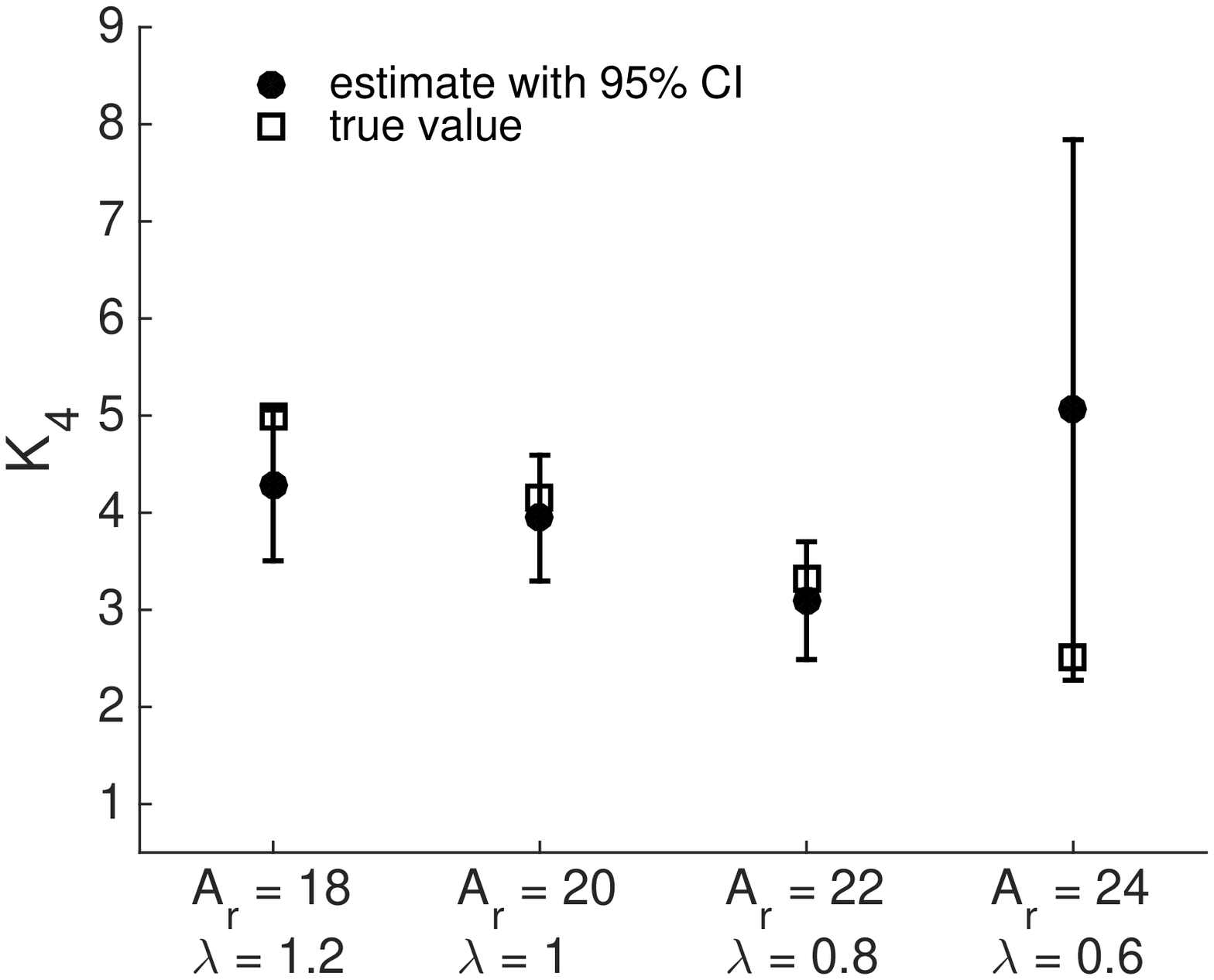}}\\
\caption*{Figure S6: \small{Comparison of best-fit parameter values and the true values for simulated data shown in Fig.\ S5. Error bars indicate 95\% CIs of the estimates. Full fitting results are given in Supplementary Table S5.}}
\end{figure}
\newpage

\begin{figure}[ht!]
\centering
\includegraphics[scale=0.7]{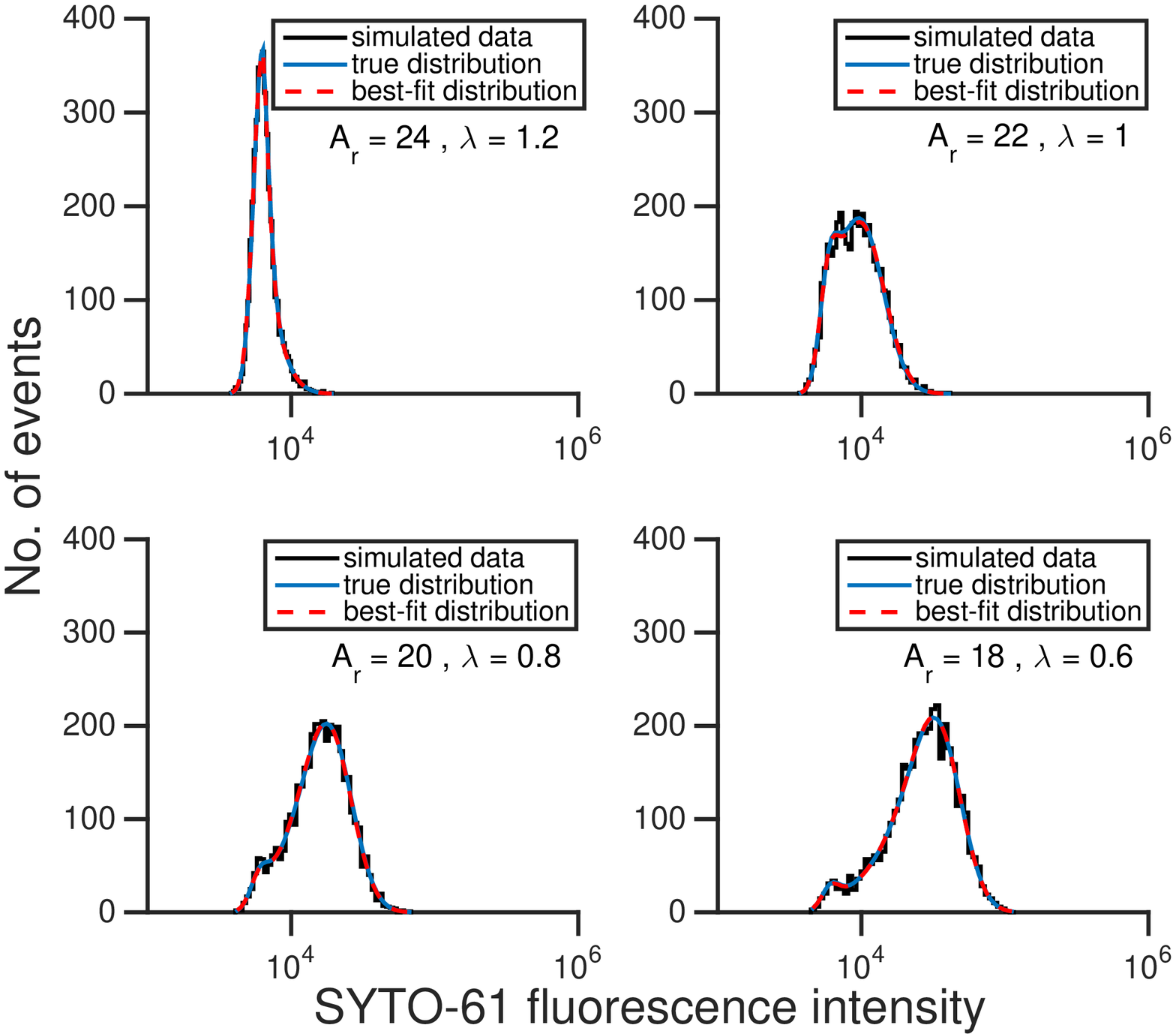}
\caption*{Figure S7: \small{Simulated data with both true distribution and best-fit distribution. 4000 samples were used for each of histograms, approximating the sample size of experimental data. The values of $A_r$ and $\lambda$ used to generate the simulated data are indicted in each panel and the other parameters are provided in Table S6.}}
\end{figure}
\newpage

\begin{figure}[ht!]
\centering
\subfloat[]{\includegraphics[scale=0.41]{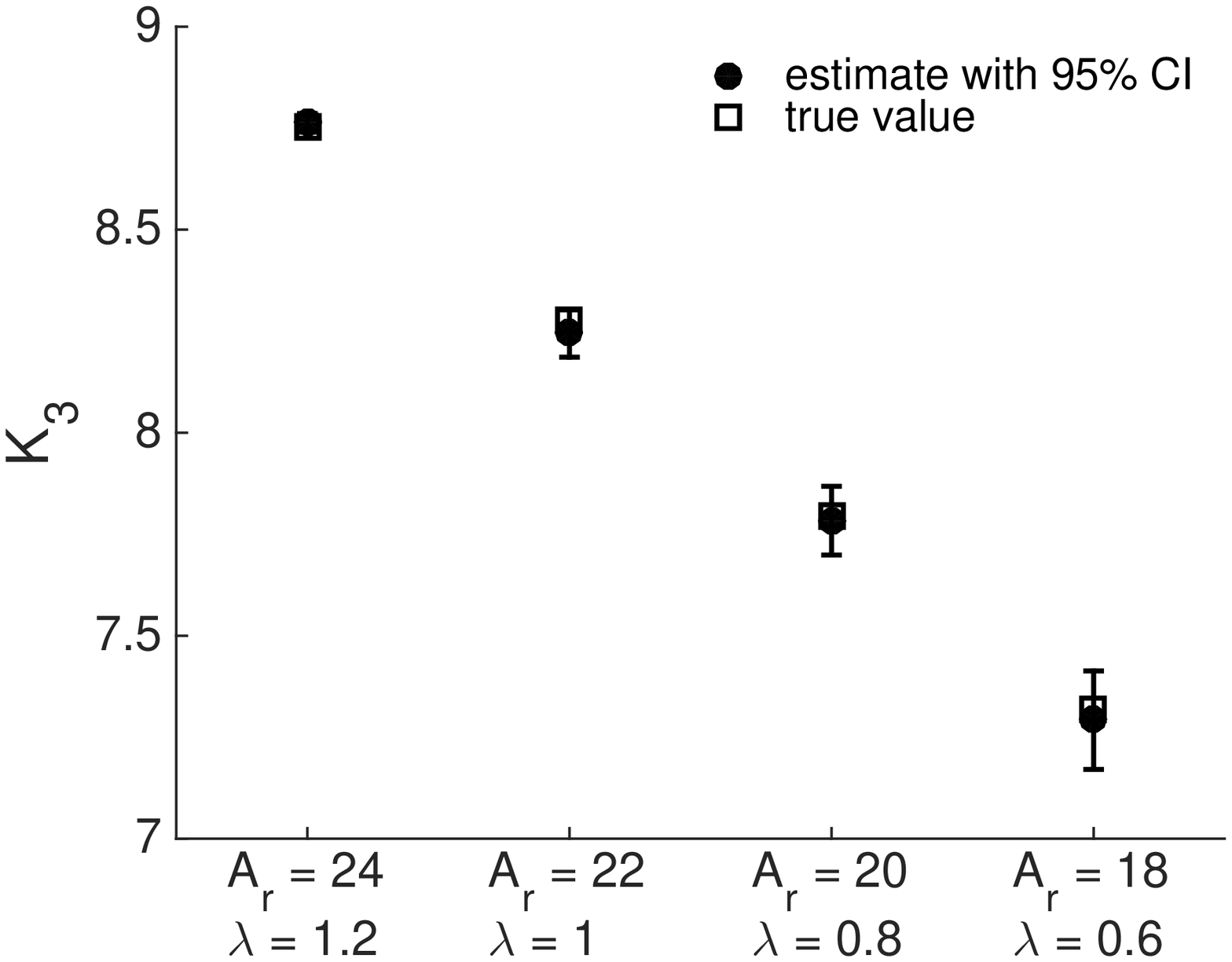}}
\subfloat[]{\includegraphics[scale=0.41]{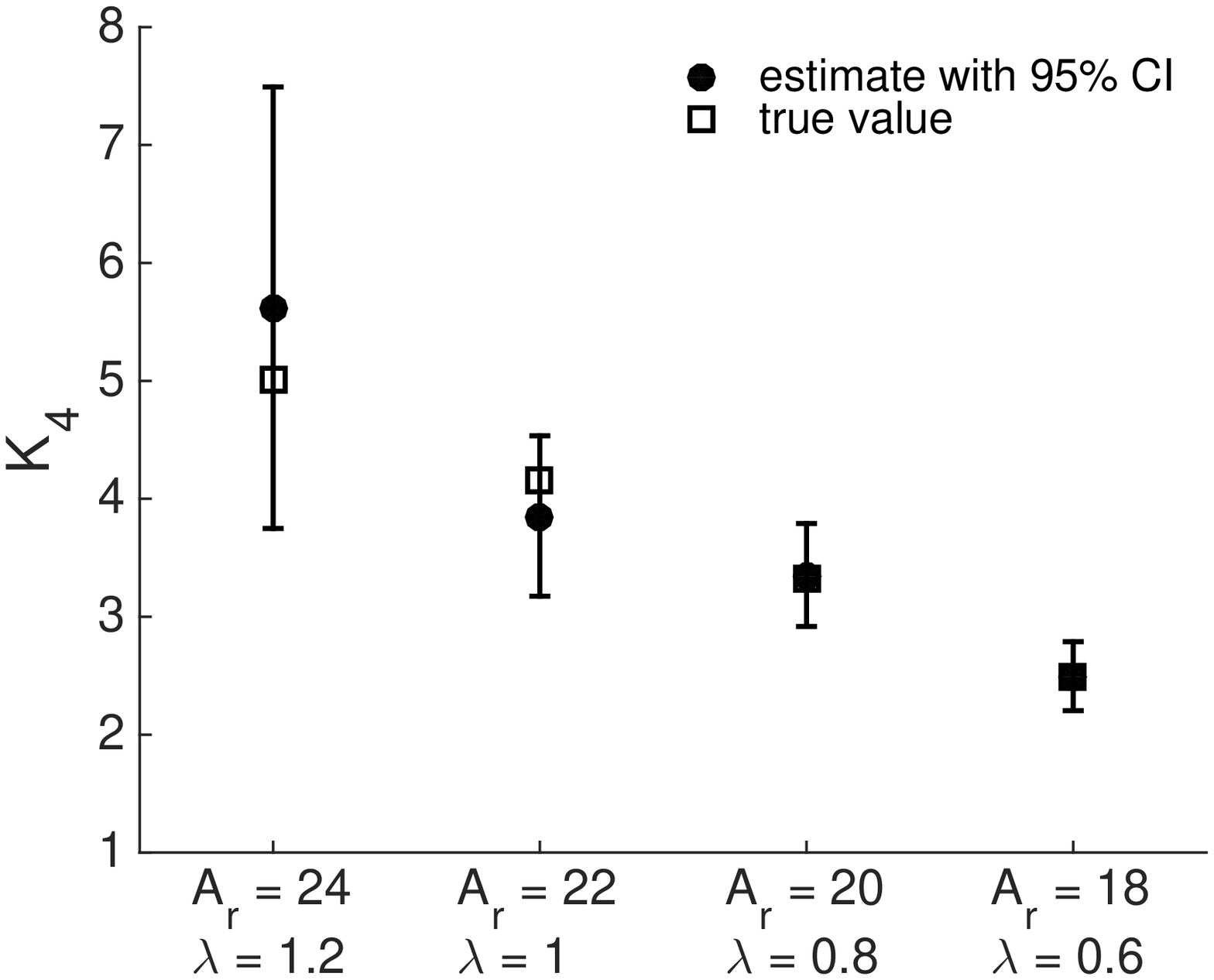}}\\
\caption*{Figure S8: \small{Comparison of best-fit parameter values and the true values for simulated data shown in Fig.\ S7. Error bars indicate 95\% CIs of the estimates. Full fitting results are given in Supplementary Table S6.}}
\end{figure}

\newpage
\subsection*{Supplementary tables}

\begin{table}[!ht]
\caption*{Table S1: \small{{{\bf The result of fitting the model to experimental histogram data (ART).} The initial point generating the best-fit parameters is given by $K_1 = 9$ (16 repeated parameters), $K_2 = 0.3$ (16 repeated parameters), $K_3 = 8$, $K_4 = 2$ and $K_5 = 2$.}}}
\begin{center}
\doublespacing
\begin{tabular}{|c|c|c|c|}
   \hline
        \multirow{2}{3cm}{ART conc. (nM)} & \multicolumn{3}{|c|}{Parameter estimate (95\% CI)}\\
\cline{2-4}
    & $K_1$ & $K_2$ & Other parameters \\
   \hline
   0 & 9.1501  (9.1232, 9.1771) & 0.3287 (0.3068, 0.3506) & \multirow{16}{3.5cm}{$K_3 = $ 8.6861 (8.6715, 8.7006); $K_4 = $ 16.6432 (11.5799, 21.7065); $K_5 = $ 1.6048 (1.5330, 1.6767).}\\
   \cline{1-3}
     0 & 9.2022 (9.1734, 9.2310) & 0.3192 (0.2978, 0.3406) & \\
   \cline{1-3}
      1.2 & 9.1868 (9.1591, 9.2145) & 0.3102 (0.2896, 0.3308) & \\
   \cline{1-3}
      2.4 & 9.1843 (9.1560, 9.2126) & 0.3293 (0.3072, 0.3514) & \\
   \cline{1-3}
      4.9 & 9.1825 (9.1549, 9.2100) & 0.3168 (0.2957, 0.3378) & \\
   \cline{1-3}
      9.8 & 9.1510 (9.1235, 9.1785) & 0.3422 (0.3197, 0.3647) & \\
    \cline{1-3}
   19.5 & 9.1355  (9.1085, 9.1625) & 0.3469 (0.3237, 0.3702) & \\
   \cline{1-3}
     39.1 & 9.0883 (9.0615, 9.1150) & 0.3598 (0.3355, 0.3840) & \\
   \cline{1-3}
      78.1 & 9.0377 (9.0118, 9.0637) & 0.3703 (0.3451, 0.3954) & \\
   \cline{1-3}
      156.3 & 9.0183 (8.9930, 9.0435) & 0.3613 (0.3372, 0.3855) & \\
   \cline{1-3}
      312.5 & 8.9805 (8.9576, 9.0034) & 0.3572 (0.3346, 0.3798) & \\
   \cline{1-3}
      625 & 8.9693 (8.9483, 8.9903) & 0.3445 (0.3232, 0.3659) & \\
   \cline{1-3}
      1250 & 8.9400 (8.9178, 8.9622) & 0.3561 (0.3338, 0.3785) & \\
   \cline{1-3}
      2500 & 8.9094 (8.8891, 8.9297) & 0.3500 (0.3291, 0.3708) & \\
   \cline{1-3}
      5000 & 8.8730 (8.8529, 8.8930) & 0.3557 (0.3352, 0.3763) & \\
    \cline{1-3}
      10000 & 8.8582 (8.8398, 8.8766) & 0.3340 (0.3151, 0.3529) & \\
   \hline
\end{tabular}
\end{center}
\end{table}

\begin{table}[!ht]
\caption*{Table S2: \small{{{\bf The result of fitting the model to experimental histogram data (DHA).} The initial point generating the best-fit parameters is given by $K_1 = 9$ (16 repeated parameters), $K_2 = 0.3$ (16 repeated parameters), $K_3 = 8$, $K_4 = 4$ and $K_5 = 2$.}}}

\begin{center}
\doublespacing
\begin{tabular}{|c|c|c|c|}
   \hline
        \multirow{2}{3cm}{DHA conc. (nM)} & \multicolumn{3}{|c|}{Parameter estimate (95\% CI)}\\
\cline{2-4}
    & $K_1$ & $K_2$ & Other parameters \\
   \hline
   0 & 9.1011  (9.0727, 9.1294) & 0.2816 (0.2599, 0.3033) & \multirow{16}{3.5cm}{$K_3 = $ 8.6243 (8.6084, 8.6402); $K_4 = $ 12.3188 (9.1899, 15.4477); $K_5 = $ 1.7590 (1.6677, 1.8502).}\\
   \cline{1-3}   
     0 & 9.1343 (9.1044, 9.1642)  &  0.2884 (0.2660, 0.3107) & \\
   \cline{1-3}
      0.12 & 9.1425 (9.1122, 9.1729)  &  0.2908 (0.2682, 0.3134) & \\
   \cline{1-3}
      0.24 & 9.1340 (9.1049, 9.1631)  &  0.2886 (0.2670, 0.3103) & \\
   \cline{1-3}
      0.5 & 9.1305 (9.1006, 9.1605)  & 0.2872  (0.2648, 0.3096) & \\
   \cline{1-3}
      1 & 9.1151 (9.0855, 9.1446)  &  0.2960 (0.2730, 0.3190) & \\
    \cline{1-3}
   2 &  9.1054 (9.0762, 9.1346)  & 0.2981 (0.2750, 0.3212) & \\
   \cline{1-3}
     3.9 & 9.0752 (9.0463, 9.1041)  & 0.3069 (0.2830, 0.3308) & \\
   \cline{1-3}
      7.8 & 9.0556 (9.0273, 9.0839)  & 0.3075 (0.2832, 0.3319) & \\
   \cline{1-3}
      15.6 & 9.0204 (8.9925, 9.0483)  & 0.3227 (0.2977, 0.3477) & \\
   \cline{1-3}
      31.3 & 8.9084 (8.8837, 8.9331)  &  0.3408 (0.3159, 0.3656) & \\
   \cline{1-3}
      62.5 & 8.8085 (8.7886, 8.8283)  &  0.3136 (0.2934, 0.3339) & \\
   \cline{1-3}
      125 & 8.6865 (8.6649, 8.7080)  &  0.3328 (0.3131, 0.3525) & \\
   \cline{1-3}
      250 & 8.6622 (8.6427, 8.6817)  &  0.3248 (0.3071, 0.3425) & \\
   \cline{1-3}
      500 & 8.6079 (8.5892, 8.6266)  &  0.3063 (0.2908, 0.3218) & \\
    \cline{1-3}
      1000 & 8.7061 (8.6912, 8.7209)  &  0.2640 (0.2491, 0.2789) & \\
   \hline
\end{tabular}
\end{center}
\end{table}

\begin{table}[!ht]
\caption*{Table S3: \small{{\bf Comparison between the fitting results and the true parameter values used to generate the simulated data shown in Fig.\ 10 in the main text.} The initial point generating the best-fit parameters is given by $K_1 = 9$, $K_2 = 0.2$, $K_3 = 8$ (4 repeated parameters), $K_4 = 2$ (4 repeated parameters) and $K_5 = 3$.}}

\begin{center}
\doublespacing
\begin{tabular}{|c||c|c||c|c|}
   \hline
       \rowcolor{LightCyan}
   Parameter & Estimate & 95\% CI & True value & Corresponding true model parameter \\
   \hline
   $K_1$ & 8.7572 & (8.7499, 8.7644) & 8.7557 & $F_0 = 20$, $r_1 = 0.24$, $\mu = 24$\\
   \hline
   $K_2$ & 0.1494 & (0.1423, 0.1564) & 0.1440 & $\sigma = 0.6$, $r_1 = 0.24$\\
   \hline
      $K_3$ & 7.9702 & (7.8422, 8.0982) & 8.0357 & $F_0 = 20$, $r_1 = 0.24$, {\color{red}$A_r = 21$}\\
   \hline
      $K_3$ & 7.9939 & (7.8759, 8.1119) & 8.0357 & $F_0 = 20$, $r_1 = 0.24$, {\color{red}$A_r = 21$}\\
   \hline
      $K_3$ & 7.9848 & (7.8660, 8.1035) & 8.0357 & $F_0 = 20$, $r_1 = 0.24$, {\color{red}$A_r = 21$}\\
   \hline
      $K_3$ & 8.0013 & (7.8823, 8.1202) & 8.0357 & $F_0 = 20$, $r_1 = 0.24$, {\color{red}$A_r = 21$}\\
   \hline
      $K_4$ & 1.0746 & (0.8548, 1.2944) & 1.2500 & $r_1 = 0.24$, {\color{red}$\lambda = 0.3$}\\
   \hline
      $K_4$ & 2.3377 & (1.8877, 2.7877) & 2.5000 & $r_1 = 0.24$, {\color{red}$\lambda = 0.6$}\\
   \hline
      $K_4$ & 3.3622 & (2.6702, 4.0542) & 3.7500 & $r_1 = 0.24$, {\color{red}$\lambda = 0.9$}\\
   \hline
      $K_4$ & 4.8077 & (3.6120, 6.0031) & 5.0000 & $r_1 = 0.24$, {\color{red}$\lambda = 1.2$}\\
   \hline
      $K_5$ & 2.2561 & (2.0861, 2.4260) & 2.3333 & $r_1 = 0.24$, $r_2 = 0.56$\\
   \hline
\end{tabular}
\end{center}
\end{table}

\begin{table}[!ht]
\caption*{Table S4: \small{{\bf Comparison between the fitting results and the true parameter values used to generate the simulated data shown in Fig.\ S3.} The initial point generating the best-fit parameters is given by $K_1 = 9$, $K_2 = 0.2$, $K_3 = 9$ (4 repeated parameters), $K_4 = 2$ (4 repeated parameters) and $K_5 = 3$.}}

\begin{center}
\doublespacing
\begin{tabular}{|c||c|c||c|c|}
   \hline
       \rowcolor{LightCyan}
   Parameter & Estimate & 95\% CI & True value & Corresponding true model parameter \\
   \hline
   $K_1$ & 8.7572 & (8.7538, 8.7606) & 8.7557 & $F_0 = 20$, $r_1 = 0.24$, $\mu = 24$\\
   \hline
   $K_2$ & 0.1452 & (0.1416, 0.1489) & 0.1440 & $\sigma = 0.6$, $r_1 = 0.24$\\
   \hline
      $K_3$ & 6.6771 & (6.5006, 6.8537) & 6.5957 & $F_0 = 20$, $r_1 = 0.24$, {\color{red}$A_r = 15$}\\
   \hline
      $K_3$ & 7.3726 & (7.2536, 7.4915) & 7.3157 & $F_0 = 20$, $r_1 = 0.24$, {\color{red}$A_r = 18$}\\
   \hline
      $K_3$ & 8.0640 & (7.9957, 8.1323) & 8.0357 & $F_0 = 20$, $r_1 = 0.24$, {\color{red}$A_r = 21$}\\
   \hline
      $K_3$ & 8.7345 & (7.9957, 8.7901) & 8.7557 & $F_0 = 20$, $r_1 = 0.24$, {\color{red}$A_r = 24$}\\
   \hline
      $K_4$ & 2.3022 & (2.0296, 2.5748) & 2.0833 & $r_1 = 0.24$, {\color{red}$\lambda = 0.5$}\\
   \hline
      $K_4$ & 2.1793 & (1.9382, 2.4203) & 2.0833 & $r_1 = 0.24$, {\color{red}$\lambda = 0.5$}\\
   \hline
      $K_4$ & 2.1407 & (1.8693, 2.4122) & 2.0833 & $r_1 = 0.24$, {\color{red}$\lambda = 0.5$}\\
   \hline
      $K_4$ & 1.8049 & (0.8657, 2.7440) & 2.0833 & $r_1 = 0.24$, {\color{red}$\lambda = 0.5$}\\
   \hline
      $K_5$ & 2.3748 & (2.2617, 2.4878) & 2.3333 & $r_1 = 0.24$, $r_2 = 0.56$\\
   \hline
\end{tabular}
\end{center}
\end{table}

\begin{table}[!ht]
\caption*{Table S5: \small{{\bf Comparison between the fitting results and the true parameter values used to generate the simulated data shown in Fig.\ S5.} The initial point generating the best-fit parameters is given by $K_1 = 9$, $K_2 = 0.2$, $K_3 = 8$ (4 repeated parameters), $K_4 = 3$ (4 repeated parameters) and $K_5 = 3$.}}

\begin{center}
\doublespacing
\begin{tabular}{|c||c|c||c|c|}
   \hline
       \rowcolor{LightCyan}
   Parameter & Estimate & 95\% CI & True value & Corresponding true model parameter \\
   \hline
   $K_1$ & 8.7549 & (8.7509, 8.7590) & 8.7557 & $F_0 = 20$, $r_1 = 0.24$, $\mu = 24$\\
   \hline
   $K_2$ & 0.1474 & (0.1442, 0.1506) & 0.1440 & $\sigma = 0.6$, $r_1 = 0.24$\\
   \hline
      $K_3$ & 7.2199 & (7.0792, 7.3606) & 7.3157 & $F_0 = 20$, $r_1 = 0.24$, {\color{red}$A_r = 18$}\\
   \hline
      $K_3$ & 7.7435 & (7.6476, 7.8395) & 7.7957 & $F_0 = 20$, $r_1 = 0.24$, {\color{red}$A_r = 20$}\\
   \hline
      $K_3$ & 8.2598 & (8.1946, 8.3250) & 8.2757 & $F_0 = 20$, $r_1 = 0.24$, {\color{red}$A_r = 22$}\\
   \hline
      $K_3$ & 8.8085 & (8.7722, 8.8448) & 8.7557 & $F_0 = 20$, $r_1 = 0.24$, {\color{red}$A_r = 24$}\\
   \hline
      $K_4$ & 4.2809 & (3.5036, 5.0583) & 5.0000 & $r_1 = 0.24$, {\color{red}$\lambda = 1.2$}\\
   \hline
      $K_4$ & 3.9442 & (3.2968, 4.5917) & 4.1667 & $r_1 = 0.24$, {\color{red}$\lambda = 1$}\\
   \hline
      $K_4$ & 3.0947 & (2.4884, 3.7009) & 3.3333 & $r_1 = 0.24$, {\color{red}$\lambda = 0.8$}\\
   \hline
      $K_4$ & 5.0574 & (2.2743, 7.8405) & 2.5000 & $r_1 = 0.24$, {\color{red}$\lambda = 0.6$}\\
   \hline
      $K_5$ & 2.2694 & (2.1635, 2.3754) & 2.3333 & $r_1 = 0.24$, $r_2 = 0.56$\\
   \hline
\end{tabular}
\end{center}
\end{table}

\begin{table}[!ht]
\caption*{Table S6: \small{{\bf Comparison between the fitting results and the true parameter values used to generate the simulated data shown in Fig.\ S7.} The initial point generating the best-fit parameters is given by $K_1 = 9$, $K_2 = 0.2$, $K_3 = 9$ (4 repeated parameters), $K_4 = 2$ (4 repeated parameters) and $K_5 = 3$.}}

\begin{center}
\doublespacing
\begin{tabular}{|c||c|c||c|c|}
   \hline
       \rowcolor{LightCyan}
   Parameter & Estimate & 95\% CI & True value & Corresponding true model parameter \\
   \hline
   $K_1$ & 8.7520 & (8.7482, 8.7558) & 8.7557 & $F_0 = 20$, $r_1 = 0.24$, $\mu = 24$\\
   \hline
   $K_2$ & 0.1458 & (0.1425, 0.1491) & 0.1440 & $\sigma = 0.6$, $r_1 = 0.24$\\
   \hline
      $K_3$ & 8.7630 & (8.7399, 8.7862) & 8.7557 & $F_0 = 20$, $r_1 = 0.24$, {\color{red}$A_r = 24$}\\
   \hline
      $K_3$ & 8.2444 & (8.1862, 8.3026) & 8.2757 & $F_0 = 20$, $r_1 = 0.24$, {\color{red}$A_r = 22$}\\
   \hline
      $K_3$ & 7.7834 & (7.6986, 7.8681) & 7.7957 & $F_0 = 20$, $r_1 = 0.24$, {\color{red}$A_r = 20$}\\
   \hline
      $K_3$ & 7.2919 & (7.1706, 7.4133) & 7.3157 & $F_0 = 20$, $r_1 = 0.24$, {\color{red}$A_r = 18$}\\
   \hline
      $K_4$ & 5.6203 & (3.7485, 7.4921) & 5.0000 & $r_1 = 0.24$, {\color{red}$\lambda = 1.2$}\\
   \hline
      $K_4$ & 3.8541 & (3.1739, 4.5343) & 4.1667 & $r_1 = 0.24$, {\color{red}$\lambda = 1$}\\
   \hline
      $K_4$ & 3.3537 & (2.9171, 3.7903) & 3.3333 & $r_1 = 0.24$, {\color{red}$\lambda = 0.8$}\\
   \hline
      $K_4$ & 2.4964 & (2.2040, 2.7889) & 2.5000 & $r_1 = 0.24$, {\color{red}$\lambda = 0.6$}\\
   \hline
      $K_5$ & 2.3139 & (2.2106, 2.4172) & 2.3333 & $r_1 = 0.24$, $r_2 = 0.56$\\
   \hline
\end{tabular}
\end{center}
\end{table}

\end{document}